\documentclass[epj-spec]{svjour}
\usepackage{graphics}
\usepackage{amssymb}
\begin{document}
\title{Colossal dielectric constants in transition-metal oxides}
\author{P. Lunkenheimer\inst{1}\fnmsep\thanks{\email{peter.lunkenheimer@physik.uni-augsburg.de}} \and S. Krohns\inst{1} \and S. Riegg\inst{2} \and S.G. Ebbinghaus\inst{3} \and A. Reller\inst{2} \and A. Loidl\inst{1}}
\institute{Experimental Physics V, Center for Electronic
Correlations and Magnetism, University of Augsburg, 86135 Augsburg,
Germany \and Solid State Chemistry, University of Augsburg, 86135
Augsburg, Germany \and Solid State Chemistry, Martin-Luther
University Halle-Wittenberg, 06120 Halle, Germany}
\abstract{Many transition-metal oxides show very large ("colossal")
magnitudes of the dielectric constant and thus have immense
potential for applications in modern microelectronics and for the
development of new capacitance-based energy-storage devices. In the
present work, we thoroughly discuss the mechanisms that can lead to
colossal values of the dielectric constant, especially emphasising
effects generated by external and internal interfaces, including
electronic phase separation. In addition, we provide a detailed
overview and discussion of the dielectric properties of
CaCu$_{3}$Ti$_{4}$O$_{12}$ and related systems, which is today's
most investigated material with colossal dielectric constant. Also a
variety of further transition-metal oxides with large dielectric
constants are treated in detail, among them the system
La$_{2-x}$Sr$_{x}$NiO$_{4}$ where electronic phase separation may
play a role in the generation of a colossal dielectric constant.
} 
\maketitle

\section{Introduction}
\label{intro}


Hot topics like high-$T_{c}$ superconductivity, colossal
magnetoresistance and multiferroicity have led to a tremendous boost
of solid state physics during the last 25 years. These and other
interesting phenomena to a large extend first have been revealed and
intensely investigated in transition-metal oxides. The complexity of
the ground states of these materials arises from strong electronic
correlations, enhanced by the interplay of spin, orbital, charge and
lattice degrees of freedom. These phenomena are a challenge for
basic research and also bear enormous potentials for future
applications as the related ground states are often accompanied by
so-called "colossal" effects, which are possible building blocks for
tomorrow's correlated electronics.

The measurement of the response of transition-metal oxides to ac
electric fields is one of the most powerful techniques to provide
detailed insight into the underlying physics that may comprise very
different phenomena, e.g., charge order, molecular or polaronic
relaxations, magnetocapacitance, hopping charge transport,
ferroelectricity or density-wave formation. The present work
concentrates on materials showing so-called colossal dielectric
constants (CDC), i.e. values of the real part of the permittivity
$\varepsilon'$ exceeding 1000. Since long, materials with high
dielectric constants are in the focus of interest, not only for
purely academic reasons but also because new high-$\varepsilon'$
materials are urgently sought after for the further development of
modern electronics. In general, for the miniaturisation of
capacitive electronic elements materials with high-$\varepsilon'$
are prerequisite. This is true not only for the common silicon-based
integrated-circuit technique but also for stand-alone capacitors.
For example, the latter, if constructed using materials with CDCs,
can reach capacitances high enough to enable their use for energy
storage, without the disadvantage of escalating volume and weight.
Such capacitors can be used, e.g., to replace batteries in hybrid
vehicles.

Most of the currently available capacitor materials with CDCs are
based on ferroelectrics, which reach very high values of the
dielectric constant often exceeding $10^{4}$. However, ferroelectric
materials exhibit a strong temperature dependence of $\varepsilon'$,
limiting their straightforward application in electronic devices.
Currently, the most prominent non-ferroelectric material showing
colossal values of $\varepsilon'$ is CaCu$_{3}$Ti$_{4}$O$_{12}$
(CCTO). Initiated by the first reports of extremely high dielectric
constants in CCTO in 2000 \cite{Subramanian2000} and further boosted
by the article of Homes \textit{et al}. \cite{Homes2001} appearing
in "Science" in the following year, until now more than 380 papers
have been published on this and related materials. The fact that
among them there are more than ten so-called "highly-cited" papers
\cite{isi} (e.g.,
\cite{Subramanian2000,Homes2001,Ramirez2000,Sinclair2002,Subramanian2002,He2002,Cohen2003,Chung2004,Lunkenheimer2002,Lunkenheimer2004,Adams2002,Raevski2003,Adams2006})
demonstrates the tremendous interest in new high-$\varepsilon'$
materials. The big advantage of CCTO compared to ferroelectric-based
dielectrics is its nearly temperature independent CDC around room
temperature. Only below about 200~K (depending on frequency) it
shows a marked and strongly frequency-dependent decrease of
$\varepsilon'(T)$ from values up to $10^{5}$ to magnitudes of the
order of 100. Many speculations about the origin of the CDCs in CCTO
have been put forward. Already in the original work by Subramanian
\textit{et al}. \cite{Subramanian2000} a barrier layer mechanism
\cite{Lunkenheimer2002,Maxwell1873,Wagner1914,Hippel1954,Ross1987}
was suggested, arising from twin boundaries. However, in some other
early works intrinsic bulk mechanisms were proposed for the
explanation of the CDCs in CCTO. This includes relaxational
excitations of highly polarizable entities of unspecified origin
\cite{Ramirez2000}, geometrically frustrated ferroelectric order
caused by symmetrical off-centre displacements of Ti ions
\cite{Homes2001} and a special kind of defects in the perovskite
structures relaxing between different equivalent configurations
\cite{Ramirez2002}. After the thorough investigation of this
material in numerous works during the last ten years, nowadays it is
quite commonly accepted that a barrier mechanism is the correct
explanation (see, e.g.,
\cite{Sinclair2002,Cohen2003,Chung2004,Lunkenheimer2002,Lunkenheimer2004,Adams2002,Bender2005,Zhang2005,Zang2005,Fiorenza2008}).
However, the nature of these barriers leading to the CDCs in CCTO
still is an open question. There are various reports on experimental
results that seem to support internal
\cite{Sinclair2002,Cohen2003,Chung2004,Adams2002,Bender2005,Zhang2005,Zang2005,Fiorenza2008}
or surface barrier layer capacitors
\cite{Lunkenheimer2004,Wang2006,Deng2007,Krohns2007,Krohns2008}
(IBLCs or SBLCs, respectively) giving rise to the detected CDCs. The
first could stem from grain boundaries (in ceramic samples) and/or
from boundaries between twins (or other planar defects) within
single crystals or within the crystallites of ceramic samples. SBLCs
may arise from the depletion layers of Schottky or
metal-insulator-semiconductor (MIS) diodes at the interfaces between
the metallic electrodes and the bulk sample.

Despite huge efforts, the properties of CCTO seem to be not ideal
for straightforward application if considering the amplitude of its
dielectric loss, the downscaling of its properties to miniaturised
devices and its applicability at frequencies in the increasingly
important frequency range around GHz
\cite{Lunkenheimer2004,Krohns2007}. Thus there is still an ongoing
search for new, better materials. There is a vast number of
materials isostructural to CCTO, which may also show CDCs but have
only partly been characterised so far
\cite{Subramanian2000,Subramanian2002,Krohns2009a,Sebald2009,Parkash2006,Parkash2008}.
But also completely different transition-metal oxides have been
reported to exhibit CDCs and it seems that this phenomenon is quite
common in this class of materials (e.g.,
\cite{Lunkenheimer2002,Samara1990,Rey1992,Mazzara1993,Chern1995,Shi1998,Chern1998,Rivas2004,Biskup2005,Cohn2005,Park2005,Ikeda2005,Liu2008,Sichelschmidt2001,Bobnar2002a,Bobnar2002b,Ritus2002,Lunkenheimer2003,Renner2004,Lunkenheimer2006,Krohns2009b,Krohns2010}).
In addition other mechanisms for the generation of CDCs are
considered. For example, the formation of CDCs by internal
interfaces spontaneously arising via electronic phase separation
(e.g., stripe-ordering) seems one of the most feasible ways to
accomplish materials that are suitable for application.

In the following we will give an overview of the different ways of
generating CDCs, laying special emphasis on interface-related
mechanisms. Then we will provide and discuss various results on
CCTO, CCTO-related materials and other transition-metal oxides with
CDCs.

\section{Mechanisms giving rise to colossal dielectric constants}
\label{sec:ldc}

The most prominent mechanisms that can give rise to enhanced values
of the dielectric constant are ferroelectricity, charge-density wave
formation, hopping charge transport, the metal-insulator transition
and various kinds of interface effects. In a first step we would
like to get a rough guess about the maximum dielectric constant that
can be reached in ionic solids without permanent dipole moments and
without invoking any of these mechanisms. The static dielectric
constant $\varepsilon_{s}$ in an ionic solid is given by
$\varepsilon_{s}=\varepsilon_{\infty}+(\Omega_{p}/\omega_{T})^{2}$.
Here $\varepsilon_{\infty}$ is the electronic dielectric constant
determined at frequencies beyond the phonon modes, $\Omega_{p}$ is
the effective ionic plasma frequency and $\omega_{T}$ the
eigenfrequency of a transverse optical vibration, which can be
directly observed in infrared experiments. The ionic plasma
frequency is given by
$\Omega_{p}^{2}=\varepsilon_{\infty}N(Ze)^{2}/(V\varepsilon_{0}\mu)$,
where $N$ is the number of ionic pairs of effective charge $Ze$ per
volume $V$, $\varepsilon_{0}$ is the vacuum dielectric permittivity
and $\mu$ the reduced mass of the ion pair involved in the
eigenmode. Large dielectric constants demand large ionic plasma
frequencies and small eigenfrequencies of the infrared active mode.
Large plasma frequencies are reached via large effective charges and
low masses. It is clear that in the oscillator model small masses
and small eigenfrequencies are related via $\omega\propto
1/\sqrt{\mu}$ and can hardly be reached at the same time if
soft-mode scenarios are not taken into consideration. To get some
estimate of the largest possible value of the dielectric constant we
make the following assumptions: $\varepsilon_{\infty}=10$, $Z=3$,
$N=1$ per unit cell volume $V=64$~{\AA}$^{3}$ and $\mu=12$ in units
of the atomic mass constant. These numbers can be reached assuming
an ion pair with Ti$^{4+}$ vibrating against O$^{2-}$ and yield an
effective ionic plasma frequency $\Omega_{p}/2\pi\approx 72$~THz.
The resulting dielectric constant now depends significantly on the
eigenfrequency of the mode under consideration. Assuming
$\omega_{T}/2\pi=2.3$~THz yields $\varepsilon_{s}\approx 1000$;
taking an eigenfrequency which better correlates with the low
reduced masses, namely $\omega_{T}/2\pi=7$~THz, results in a
dielectric constant $\varepsilon_{s}\approx 120$, which comes close
to the maximum value of experimentally observed dielectric constants
in purely ionic solids. Thus for the generation of CDCs, other
mechanisms have to be considered as discussed in the following.

\subsection{Ferroelectricity}

It is known since long that in ferroelectrics very high values of
the dielectric constant can arise \cite{Halblutzel1939,Lines1979}.
Approaching the ferroelectric phase transition at $T_{c}$ from high
temperatures, $\varepsilon'(T)$ strongly increases, usually
following a Curie-Weiss behaviour and starts to decrease again below
$T_{c}$. In addition, ferroelectrics have pronounced non-linear
dielectric properties, e.g., showing characteristic hysteresis loops
of the electric-field dependent polarisation \cite{Lines1979}. Both
phenomena represent problems for application in electronic devices.
This partly can be overcome by doping and special processing thereby
adjusting microstructure and internal interfaces. The well-known
ceramic Barrier Layer Capacitors use a combination of interface
polarisation effects and ferroelectric materials like BaTiO$_{3}$ to
achieve high capacitance values with temperature and voltage
dependences that are tolerable at least for some applications
\cite{Fujimoto1985,Yang1996,Moulson1990}. Ferroelectric transitions
often are classified as displacive or order-disorder type. The
latter case corresponds to the ordering of dipolar degrees of
freedom already present at $T>T_{c}$. In these systems the hopping
of the dipoles can lead to strong frequency dependence of
$\varepsilon'$ at technically relevant frequencies (Hz-GHz) making
them less suited for application
\cite{Lines1979,Staresinic2006,Schrettle2009}. In contrast,
ferroelectrics of displacive type usually show no frequency
dependence up to infrared frequencies, where the well-known
soft-phonon modes appear \cite{Lines1979}.

A special variant of ferroelectrics are the so-called relaxor
ferroelectrics \cite{Cross1987,Bokov2006}. Their static dielectric
constant shows a strong increase with decreasing temperature just as
for canonical ferroelectrics. However, this is superimposed by a
marked relaxation mode that leads to peaks in $\varepsilon'(T)$ at
temperatures that are strongly dependent on frequency. Different
explanations have been proposed for this behaviour, e.g., in terms
of polar cluster dynamics, but no consensus has been achieved so far
\cite{Cross1987,Bokov2006,Viehland1990,Westphal1992,Vugmeister1998}.

While in conventional and relaxor ferroelectrics ions or dipoles are
the relevant entities achieving ferroelectric order, also the
ordering of electronic degrees of freedom has been considered
\cite{Ikeda2005,Portengen1996,Batista2002,Kampf2003}. For example,
the occurrence of CDCs of magnitude $>4000$ detected in the
mixed-valent transition-metal oxide LuFe$_{2}$O$_{4}$ was ascribed
to an electronic polarisation mechanism involving charge ordering of
Fe$^{2+}$ and Fe$^{3+}$ ions \cite{Ikeda2005}. Also in certain
charge-transfer salts a ferroelectric transition of electronic
origin recently was discussed \cite{Staresinic2006,Monceau2001}.

Finally, it should be mentioned that in some materials, the
so-called incipient ferroelectrics, long-range ferroelectric order
is prevented by quantum fluctuations at low temperature, setting in
before complete order is achieved at $T_{c}$. At temperatures
sufficiently above $T_{c}$, $\varepsilon'(T)$ of these materials
follows the Curie-Weiss law and thus CDCs are observed. The most
prominent incipient ferroelectric is SrTiO$_{3}$ \cite{Muller1979},
which shows a tendency of $\varepsilon'(T)$ to saturate at low
temperatures, setting in below about 30~K due to the mentioned
quantum effects \cite{Viana1994}. Besides BaTiO$_{3}$, also
SrTiO$_{3}$ is often employed as dielectric material in commercial
ceramic Barrier Layer Capacitors.

\subsection{Charge-Density Waves}

In highly anisotropic low-dimensional materials a metal-insulator
transition can arise with lowering of the temperature, which is
accompanied by the formation of a charge-density wave (CDW). Here
the electronic charge density is a periodic function of position and
its period can be incommensurate with the crystal lattice. A very
well-known example of a CDW system is also found in the group of
transition-metal oxides, namely the blue bronze, K$_{0.30}$MoO$_{3}$
\cite{Dumas1983}. The dielectric behaviour of CDW systems shows two
characteristic features: A harmonic oscillator mode at GHz
frequencies caused by the CDW being pinned at defects and a huge
relaxation mode at kHz-MHz involving colossal values of the
dielectric constant \cite{Gruner1988,Cava1986,Cava1985}. In CDW
systems the highest intrinsic dielectric constants of any materials
are observed, reaching magnitudes of up to $10^{8}$. Littlewood
\cite{Littlewood1987} has proposed screening effects of the pinned
CDW by the normal electrons, not participating in the CDW, to
explain the occurrence of the low-frequency relaxation mode and CDCs
in this class of materials. Due to the strong frequency dependence
of $\varepsilon'$ and the high dielectric losses associated with the
relaxational modes, CDWs are not applied in capacitive devices.

\subsection{Hopping charge transport}
\label{sec:hop}

 Hopping conductivity is the most common charge-transport process in condensed matter. As it is intimately related
with the occurrence of a power law with negative exponent in the
frequency-dependent dielectric constant $\varepsilon'(\nu)$, it will
always lead to a divergence of $\varepsilon'$ for low frequencies
\cite{Long1982,Elliott1987,Jonscher1977,Jonscher1983}. Hopping
conduction is the typical charge-transport process of localised
charge carriers. In electronic conductors, electrons (or holes) can
localise due to disorder. Disorder may arise from an amorphous
structure, from doping (substitutional disorder) or occur even in
nominally pure crystals due to slight deviations from stoichiometry
or lattice imperfections. Hopping conduction leads to a
characteristic signature in the frequency dependence of the complex
conductivity, namely a power-law increase $\sigma'=\sigma_{0}
\nu^{s}$ with the exponent $s<1$
\cite{Long1982,Elliott1987,Mott1979}. This power law was shown by
Jonscher \cite{Jonscher1977,Jonscher1983} to be a quite universal
phenomenon in all types of disordered matter and termed "Universal
Dielectric Response" (UDR). This behaviour can be understood in the
framework of various models on the charge transport of localised
charge carriers, including the often-employed variable-range hopping
(VRH) model \cite{Long1982,Elliott1987,Mott1979}. These models
originally were developed for amorphous and highly doped
semiconductors like doped silicon but also, e.g., for thin
scandium-oxide films
\cite{Long1982,Elliott1987,Mott1979,Pollak1961,Pike1972}. The
typical signature of hopping transport in measurements of the ac
conductivity was also found in numerous transition-metal oxides
(e.g.,
\cite{Lunkenheimer2004,Krohns2007,Sichelschmidt2001,Bobnar2002a,Bobnar2002b,Lunkenheimer2003,Renner2004,Lunkenheimer2006,Krohns2009b,Pike1972,Lunkenheimer1991,Lunkenheimer1992,Seeger1999}).
Via the Kramers-Kronig relation, the $\nu^{s}$ power law also leads
to a corresponding power law in the imaginary part of the ac
conductivity, namely $\sigma''=\tan(s\pi/2)\sigma_{0}\nu^{s}$
\cite{Jonscher1983}. As the dielectric constant is directly related
to $\sigma''$ via $\varepsilon'=\sigma''/(2\pi\nu\varepsilon_{0})$
(with $\varepsilon_{0}$ the permittivity of vacuum) hopping
conduction is expected to lead to a power law $\varepsilon'\propto
\nu^{s-1}$. Thus, as $s<1$, the dielectric constant can easily reach
colossal magnitudes for low frequencies. However, as the factor
$tan(s\pi/2)$ usually is of the order of one, the dielectric loss
$\varepsilon''=\sigma'/(2\pi\nu\varepsilon_{0})$ is relatively high
(which of course is reasonable for a conducting material) rendering
this effect unsuited for application.

\subsection{Metal-insulator transition}

It is known since about 100 years that the Clausius-Mosotti relation
will lead to a polarisation catastrophe, i.e. a divergence of the
dielectric constant when approaching the metal-insulator (MI)
transition from the insulating side \cite{Castner1980}. It is
naively clear that the reduction of the restoring forces experienced
by electrons localised at atomic sites, which should occur when the
material approaches the metallic state with its itinerant electron
states, will lead to an increase of the electronic polarisability
and thus the dielectric constant. Indeed, such a divergence has been
observed in some cases
\cite{Castner1980,Capizzi1980,Hess1982,Pimenov1999,Kundys2006}, the
most prominent one being measurements of the dielectric properties
of doped silicon for increasing doping level \cite{Hess1982}.
Several theoretical approaches have appeared treating this topic and
going beyond the simple arguments based on the Clausius-Mosotti
equation \cite{Mott1990,McMillan1981,Bhatt1984,Aebischer2001}.
Prototypical MI transitions are regularly found in transition-metal
oxides, the most well-known ones being those in magnetite,
Fe$_{3}$O$_{4}$, and vanadium oxide, V$_{2}$O$_{3}$. One may
speculate if some of the observations of CDCs in transition-metal
oxides
\cite{Lunkenheimer2002,Samara1990,Rey1992,Mazzara1993,Chern1995,Shi1998,Chern1998,Rivas2004,Biskup2005,Cohn2005,Park2005,Ikeda2005,Liu2008,Sichelschmidt2001,Bobnar2002a,Bobnar2002b,Ritus2002,Lunkenheimer2003,Renner2004,Lunkenheimer2006,Krohns2009b,Kundys2006}
could be due to the fact that these materials often are at the verge
of a MI transition. However, in most cases interfacial effects as
treated in the following section seem the more likely explanation.
The expected divergence of $\varepsilon'$ at the MI transition of
course is also accompanied by an increase of the conductivity,
preventing any technical application of this effect for the
construction of capacitive components.

\subsection{Interface effects}
\label{sec:interface}

 Interfaces of any kind can generate very high
apparent values of the dielectric constant because they act as
parallel-plate capacitors with very small plate distances, thus
having high capacitances \cite{Lunkenheimer2002}. This corresponds
to the long-known Maxwell-Wagner (MW) polarisation effects.
Originally, Maxwell and Wagner developed their models for three
different types of heterogeneous samples, namely for a dielectric
sample consisting of layers with different dielectric properties
(i.e. $\varepsilon'$ and $\sigma'$) arranged perpendicular to the
electric field, for an ideal dielectric sample covered with a bad
conductor of varying thickness and for a homogeneous dielectric
medium in which spherical particles with different dielectric
properties are suspended \cite{Maxwell1873,Wagner1914}. For all
these cases, strong dispersion of the dielectric constant and of the
loss was deduced, which can be completely understood from the
heterogeneity of the investigated samples without invoking any
frequency-dependent microscopic processes within the dielectric
materials.

\begin{figure}[htb] \centerline{\resizebox{0.4\columnwidth}{!}{
\includegraphics{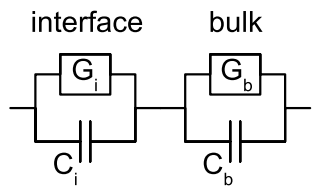}}}
\caption{Equivalent circuit representing the simplest model for a
heterogeneous sample, composed, e.g., of an interfacial and bulk
region.} \label{fig:circuit}
\end{figure}

A straightforward approach to understand the dielectric behaviour of
such a heterogeneous system is an equivalent circuit analysis
\cite{Ross1987,Jonscher1983}. Let us assume the simplest case of the
sample being composed of two regions with different dielectric
properties. Each region, of course, has some conductivity and
dielectric constant, i.e. represents a lossy capacitor. It shall be
modeled by a simple parallel RC circuit, i.e. we assume that there
is no intrinsic frequency dependence of conductivity or dielectric
constant. For the cases treated by Maxwell and Wagner
\cite{Maxwell1873,Wagner1914} and for any cases where the second
dielectric region is of interfacial type as, e.g., a grain boundary
or a surface depletion layer, it seems reasonable to assume a serial
connection of the two RC circuits as depicted in Fig.
\ref{fig:circuit}. Then the total admittance of this circuit is
given by $Y_{t}= G'_{t}+iG''_{t}=(G_{i}+i\omega C_{i})(G_{b}+i\omega
C_{b})/[(G_{i}+i\omega C_{i})+(G_{b}+i\omega C_{b})$]
($\omega=2\pi\nu$ is the circular frequency). When resolved into
real and imaginary part and calculating the capacitances via
$C_{t}'=G''/\omega$ and $C_{t}''=G'/\omega$, we obtain:

\begin{equation}
C_{t}^{\prime \prime }=\frac{G_{i}G_{b}(G_{i}+G_{b})+\omega
^{2}(G_{i}C_{b}^{2}+G_{b}C_{i}^{2})}{\omega (G_{i}+G_{b})^{2}+\omega
^{3}(C_{i}+C_{b})^{2}} \label{eq:Ct1}
\end{equation}

\begin{equation}
C_{t}^{\prime }=\frac{(G_{i}^{2}C_{b}+G_{b}^{2}C_{i})+\omega
^{2}C_{i}C_{b}(C_{i}+C_{b})}{(G_{i}+G_{b})^{2}+\omega
^{2}(C_{i}+C_{b})^{2}} \label{eq:Ct2}
\end{equation}

\begin{figure}[tbp] \centerline{\resizebox{0.5\columnwidth}{!}{
\includegraphics{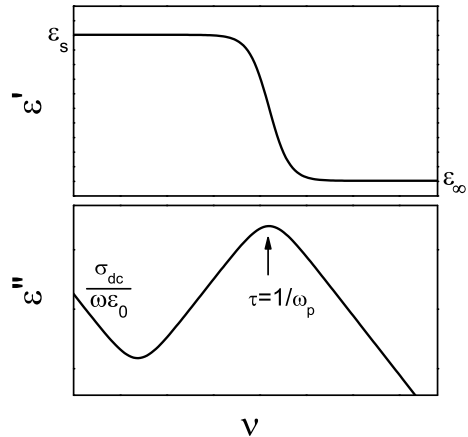}}}
\caption{Dielectric response for a Debye relaxation arising from the
polarisation of dipolar degrees of freedom with an additional
contribution from dc charge transport. In the loss a peak arises in $\omega_p = 1/\tau$. Identical spectra are also
produced by the equivalent circuit shown in Fig. \ref{fig:circuit}
without invoking any intrinsic frequency dependence of the circuit
elements.} \label{fig:debye}
\end{figure}

\noindent Here $G_{b}$ and $G_{i}$ are the conductances and $C_{b}$
and $C_{i}$ are the capacitances of the different sample regions
with indices $b$ and $i$ standing for bulk and interface,
respectively. If neglecting the geometry factors connecting $C$ and
$\varepsilon'$, these relations lead to exactly the same frequency
dependence as the Debye relaxation laws,

\begin{equation}
\varepsilon'=\varepsilon_{\infty}+\frac{\varepsilon_{s}-\varepsilon_{\infty}}{1+\omega^{2}\tau^{2}}
\label{eq:Debe1}
\end{equation}

\noindent  and

\begin{equation}
\varepsilon''=\frac{(\varepsilon_{s}-\varepsilon_{\infty})\omega\tau}{1+\omega^{2}\tau^{2}}
+ \frac{\sigma_{dc}}{\omega\varepsilon_{0}},
\label{eq:Debe2}
\end{equation}

\noindent describing the relaxational response of an ideal dipolar
system (Fig. \ref{fig:debye}) \cite{Kremer2002}. Here
$\varepsilon_{s}$ and $\varepsilon_{\infty}$ are the low and high
frequency limits of $\varepsilon'(\omega)$, respectively, and $\tau$
is the relaxation time, describing the dipolar dynamics. The last
term in Eq. (\ref{eq:Debe2}) accounts for the contribution of dc
charge transport to the loss, with $\sigma_{dc}$ the dc
conductivity. The quantities of Eqs. \ref{eq:Ct1} and \ref{eq:Ct2}
corresponding to $\varepsilon_{s}$, $\varepsilon_{\infty}$ and
$\tau$ are:

\begin{equation}
C'_{s}=\frac{G_{i}^{2}C_{b}+G_{b}^{2}C_{i}}{(G_{i}+G_{b})^{2}}
\label{eq:Cs}\end{equation}

\begin{equation}
C'_{\infty}=\frac{C_{i}C_{b}}{C_{i}+C_{b}}\end{equation}

\begin{equation}
\tau=\frac{C_{i}+C_{b}}{G_{i}+G_{b}}\end{equation}

\noindent Nowadays any relaxational response generated by
heterogeneities in the sample is termed MW relaxation.

Let us now assume that one of the two dielectric regions of the
sample indeed is an insulating interface like the depletion layer of
a Schottky diode at the electrodes or like barrier layers between
the grains of a polycrystalline sample. In the latter case we make
the simplifying assumption that all grain boundaries act like a
single capacitor, which is justified in most cases (there are more
detailed treatments like the brick layer model
\cite{Ross1987,VanDijk1981,Verkerk1982} but they lead to essentially
the same results). Then we can assume that $C_{i} \gg C_{b}$ (the
interface is much thinner than the bulk) and that $G_{i} \ll G_{b}$
(the interface is nearly insulating). In this limit we get
$C'_{s}=C_{i}$, $C'_{\infty}=C_{b}$ and $\tau=C_{i}/G_{b}$. Thus at
low frequencies, the dielectric constant is completely governed by
the large capacitance of the thin insulating interface layer and
only at high frequencies the intrinsic bulk response is detected (by
the way, the same goes for the conductance, i.e.
$G'_{t}(\nu\rightarrow 0)=G_{i}$ and $G'_{t}(\nu\rightarrow
\infty)=G_{b}$). This can be rationalised by the bridging of the
high contact resistance by the contact capacitance, acting like a
short at high frequencies \cite{Lunkenheimer2002}. When evaluating
dielectric measurements, the dielectric constant usually is
calculated from the measured capacitance data via
$\varepsilon'=C'/C_{0}$. $C_{0}$ is the geometrical capacitance, for
a parallel-plate capacitor given by $C_{0}=\varepsilon_{0} A/t$ with
$A$ the area and $t$ the plate distance of the capacitor. At low
frequencies, in the interface dominated regime, of course $C_{0}$
determined from the bulk geometry is the wrong quantity and can be
many decades smaller than the geometrical capacitance of the thin
insulating layer(s). Thus, despite the true dielectric constant of
the interface region (and of course that of the bulk) have "normal"
magnitudes, say of the order of ten, very large apparent values of
$\varepsilon'$ can arise \cite{Lunkenheimer2002}. It is difficult to
distinguish these effects from true bulk contributions as the
resulting relaxation modes have all the characteristics of intrinsic
relaxations, where dipolar entities are assumed to reorient in
accord with the ac field at low frequencies but to be unable to
follow its quick variations at high frequency. Also the typical
frequency-dependent shifts of the relaxation features, mirroring in
the intrinsic case the freezing of dipolar dynamics at low
temperatures \cite{Kremer2002}, are observed: The bulk conductance
$G_{b}$ of the considered materials usually has semiconducting
temperature characteristics, i.e. it increases exponentially with
temperature. Thus, via the relation $\tau=C_{i}/G_{b}$ the
relaxation time will exhibit the typical strong temperature
variation of dipolar systems leading to pronounced shifts of the
step in $\varepsilon'(\nu)$ and the peak in $\varepsilon''(\nu)$ to
low frequencies when temperature is reduced, just as for an
intrinsic dipolar relaxation.

\begin{figure}[htb]
\centerline{\resizebox{0.65\columnwidth}{!}{
  \includegraphics{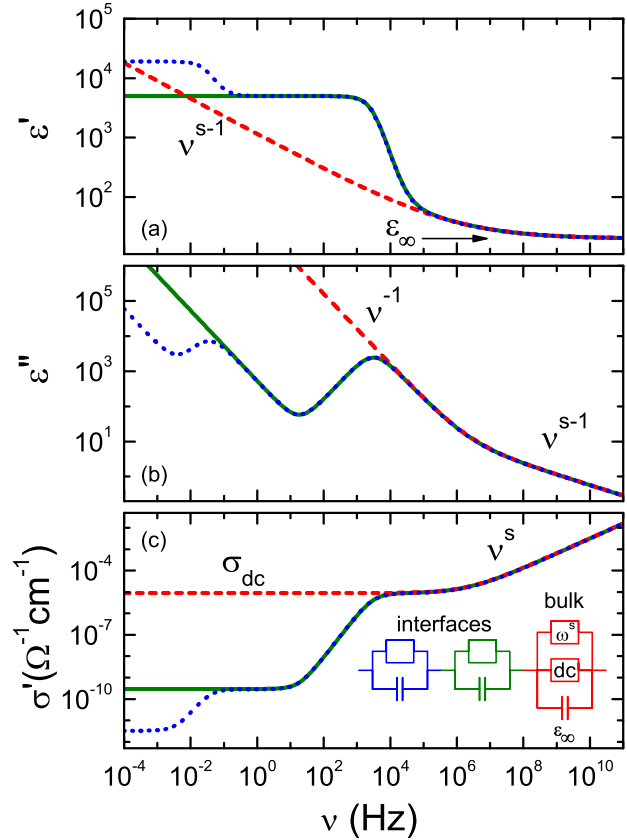} }}
\caption{Frequency-dependent dielectric response for the equivalent
circuit shown in (c). Dashed lines: Intrinsic bulk response. Solid
and dotted lines: Overall response with one, respectively two
additional RC-circuits. The circuit parameters have been chosen to
reveal the prototypical behaviour of doped semiconductors with
Schottky-barrier-type contacts \cite{Lunkenheimer2002}.}
\label{fig:theo}
\end{figure}

Real-life samples exhibiting interfacial polarisation usually do not
exactly follow the behaviour suggested by Eqs. \ref{eq:Ct1} and
\ref{eq:Ct2} and depicted in Fig. \ref{fig:debye}. For example, the
step in $\varepsilon'(\nu)$ and the peak in $\varepsilon''(\nu)$
sometimes are smeared out and significantly broader than expected.
Just as for systems showing intrinsic dipolar relaxations
\cite{Sillescu1999,Ediger2000}, this can be ascribed to a
distribution of relaxation times. For heterogeneity-generated MW
relaxations it can be assumed to arise, e.g., from the roughness of
the surface in case of a Schottky-diode mechanism or from the
distribution of grain (and thus also grain-boundary) sizes in
ceramic samples. In addition, the ubiquitous hopping conductivity
usually leads to the typical power-law frequency dependences of the
UDR \cite{Jonscher1983}, i.e. $\sigma'\propto \nu^{s}$,
$\sigma''\propto \nu^{s}$, $\varepsilon'\propto \nu^{s-1}$ and
$\varepsilon''\propto \nu^{s-1}$ with $s<1$ (see Sec.
\ref{sec:hop}), which determines the spectra in the high-frequency
regime where the intrinsic bulk behaviour governs the dielectric
response. The resulting curves are schematically shown in Fig.
\ref{fig:theo} (solid lines) \cite{Lunkenheimer2002}. The dashed
lines represent the pure intrinsic behaviour of the bulk sample,
which would be observed in the absence of any interface
contributions. Finally, in heterogeneity dominated samples often an
additional relaxation leading to even higher values of
$\varepsilon'$ at the lowest frequencies is found as schematically
illustrated by the dotted lines in Fig. \ref{fig:theo}. This implies
the presence of two different interfacial regions in the sample,
which can be rationalised, e.g., by the simultaneous presence of a
surface depletion layer and insulating grain or twinning boundaries
\cite{Lunkenheimer2004,Krohns2007,Krohns2008}. Thus, the most general
equivalent circuit is provided by two parallel RC circuits,
connected in series to the bulk element, the latter  consisting of a
capacitance, dc resistor and a frequency-dependent complex
conductance element accounting for the hopping-generated UDR (inset
of Fig. \ref{fig:theo}(c)).

As mentioned above, insulating interfaces can occur at the surface
of the sample, e.g., due to the formation of a Schottky diode at the
electrode/sample contact or due to a thin insulating surface layer
with slightly different stoichiometry (e.g., oxygen content). Also
internal interfaces can arise, e.g., from grain boundaries in
ceramic samples or planar crystal defects (e.g., twin boundaries).
In general, surface-related effects (SBLCs) represent a problem for
technical application as they do not scale down in accord with
sample size and especially in the common thin-film techniques
employed in integrated electronics, much smaller CDCs, if colossal
at all, may arise. One may run into similar problems for
grain-boundary or defect-generated IBLCs, at least if the film
thickness approaches the spatial dimensions of the regions separated
by these types of interfaces (grains or crystallites). Nevertheless,
both kinds of IBLCs are used in some types of stand-alone
high-capacitance condensers as the mentioned ceramic barrier-layer
capacitors and so-called double-layer capacitors (also termed
supercapacitors) \cite{Simon2008}. The latter can reach capacitances
up to several kF making them suitable for energy storage instead of
batteries. They use the well-known blocking-electrode effect
\cite{Ross1987} observed at low frequencies in ionically conducting
materials. In contrast to electrons or holes in electronic
conductors, ionic charge carriers cannot penetrate into the metallic
electrodes used to contact the samples and instead accumulate in
thin layers immediately below the sample surface. These insulating
and very thin layers represent large capacitors leading, by means of
the equivalent circuit discussed above, to an enhanced capacitance
of the sample. For both supercapacitors and ceramic barrier layer
capacitors, special preparation techniques are necessary for
optimisation of the microstructure. For example in supercapacitors,
electrode materials with large inner surfaces like activated
charcoal are commonly used.

Of high interest for application may also be any spontaneously
arising interfaces caused, e.g., by electronic phase separation or
charge order. These phenomena were found in many transition-metal
oxides as the colossal magnetoresistance manganites
\cite{Yamada1996,Uehara1999,Moreo1999,Dagotto2001} or the cuprate
superconductors
\cite{Emery1993,Tranquada1995,Tranquada2004,Reznik2006}. Also the
system La$_{2-x}$Sr$_{x}$NiO$_{4}$ attracted much interest due to
the formation of stripe-like electronic phase separation in large
parts of its phase diagram
\cite{Chen1993,Sachan1995,Lee1997,Tranquada1996,Yamanouchi1999,Du2000,Lee2002,Tranquada2002,Ishizaka2004}.
In the charge-ordered or phase-separated phases of many materials, a
large or even colossal magnitude of the dielectric constant has been
detected (e.g.,
\cite{Rivas2004,Park2005,Liu2008,Krohns2009b,Krohns2010,Rivadulla1999,Mercone2004,Mira2006,Filippi2009,Garcia2009}).
The true reason for the found high values of $\varepsilon'$ often is
not completely clarified but it seems well possible that at least in
some cases internal interfaces between the different phase-separated
regions may play a role \cite{Krohns2009b,Mercone2004}. An enhanced
dielectric constant due to interfaces between these spontaneously
forming heterogeneous regions within the sample still represents a
kind of MW effect and no intrinsically large $\varepsilon'$ in any
parts of the sample has to be assumed to explain the CDC. However,
within this scenario the MW relaxation leading to the apparently
colossal $\varepsilon'$ would arise from an effect that is inherent
to the material and does not depend on sample preparation and thus
the observed CDCs could be regarded as quasi-intrinsic. In addition,
in this case the heterogeneity arises on a much finer scale than,
e.g., for contact or grain boundary effects and can be expected to
persist also for the scaled down dimensions prevailing in integrated
electronics.

\subsection{Distinguishing CDCs from interface and intrinsic bulk effects}
\label{sec:distin}

Distinguishing MW relaxations from intrinsic bulk ones is not a
trivial task. As mentioned above, both phenomena can lead to
absolutely identical frequency and temperature characteristics of
the dielectric quantities. In general, if a strong relaxation mode
is observed, with absolute values of $\varepsilon'_{s}$ exceeding
100 and without the typical temperature characteristics of
ferroelectrics or relaxor ferroelectrics, a non-intrinsic mechanism
is very likely. In such case no exotic mechanisms should be invoked
to explain the experimental observations before some additional
checks have been performed. Checking for SBLC effects is a
relatively simple task. If the formation of Schottky diodes at the
electrode-sample interface are suspected, two measurements of the
same sample with contacts formed by a conducting paste or paint
(e.g., silver paint) and by sputtering or evaporating a metal layer
should be performed. Schottky diodes only form if metal and
semiconductor are in direct contact. Conducting pastes or paints
typically contain metallic grains of $\mathrm{\mu}$m dimension. It
is easy to imagine that these grains "touch" the sample surface only
at distinct points and the "wetting" of the sample surface by the
metal is poor. Quite in contrast, sputtered or evaporated contacts
provide a much better wetting and thus a more effective formation of
the thin depletion layer of the Schottky diodes. It should be noted
that in case of silver paint at areas, where no direct contact
between metal and sample is achieved, the depletion layer of the
diode is replaced by air gaps. These gaps of course also act as
capacitors, which, however, are of much smaller magnitude due to the
larger gap width compared with the thickness of the diode depletion
layers and the low $\varepsilon'\approx 1$ of air. (If no Schottky
diodes are formed, these air-gap capacitors obviously have too small
capacitances to provide any contribution to the measured
$\varepsilon'$ as, e.g., in BaTiO$_{3}$ silver paint measurements
provide the correct values \cite{Lunkenheimer2004}.) Overall, the
values of $\varepsilon_{s}$ can be expected to be strongly reduced
when using metallic paints or pastes as electrodes. Such a finding
clearly points to diode formation at the surface causing the
observed CDCs. The thickness of Schottky-diode generated depletion
layers depends on the difference of the work functions of metal and
semiconductor. Thus at first glance it may seem feasible to simply
use two different metals without any further variation of the type
of contact. However, the work functions of typical electrode metals
do not strongly differ
(typically by the order of $10\verb"%"$ only) and the effect is barely visible, having
in mind that $\varepsilon'(T,\nu)$ often varies by several decades.
In contrast, the effect of varying the contact type usually is much
bigger and often of the order of one decade or more
\cite{Lunkenheimer2004,Krohns2007,Renner2004,Krohns2009b}.

Another origin of SBLCs could be deviations from bulk stoichiometry
at the surface, e.g., of the oxygen content that may be different to
that of the bulk due to the contact with ambient atmosphere. This
can lead to the direct generation of a contact capacitance by this
layer. Also the formation of a MIS diode with a corresponding thin
depletion layer seems possible, with the insulator being the
mentioned surface layer of different stoichiometry
\cite{Deng2007,Krohns2008,Deng2009}. In the latter case, again a variation of contact
type should lead to a marked variation of $\varepsilon_{s}$. In the
former case, the same should be achieved by polishing the sample to
remove the insulating surface layer and applying the same type of
contacts. When doing this, some care should be taken to ensure the
same surface roughness before and after polishing as it can be
essential for the formation of diodes (see above). Finally, the
presence of surface-related effects can also be simply checked for
by varying the thickness of the sample \cite{Lunkenheimer2004}. For
any kind of SBLC, the sample thickness should directly scale with
the absolute value of the CDCs because the surface capacitance
remains the same but the geometrical capacitance $C_{0}$ used to
calculate $\varepsilon'$ varies with thickness (see Sec.
\ref{sec:interface}). Thus thicker samples should have higher CDC
values. Again, when polishing down the sample, the same surface
roughness as before must be ensured.

Checking for IBLCs often is a more difficult task. The simplest way
to exclude grain boundaries is measuring single-crystalline samples.
Alternatively, different grinding of the powders before pressing and
sintering and different sintering conditions should be applied to
achieve different grain sizes. Insulating grain boundaries may arise
from deviations of stoichiometry at the grain surfaces from the bulk
one \cite{Adams2002} and varying the conditions during preparation
may lead to different conductivities, thicknesses etc. of these
surface layers. The most commonly applied way to check for
grain-boundary effects is the variation of the sintering time
applied to the final pressed sample tablets (e.g.,
\cite{Adams2002,Bender2005,Zang2005}). Usually longer sintering
times lead to larger grain sizes. The concomitant reduction of the
thickness of the internal barriers, if averaged over the whole
sample, leads to an increase of the overall capacitance of the
IBLCs. However, one should be aware that longer sintering also
should have an effect on the surface of the sample, reducing, e.g.,
its roughness due to grain growth at the surface. This may also
influence possible IBLC formation and makes the interpretation of
such experiments somewhat ambiguous. In case of single-crystalline
samples, the presence of possible IBLCs caused by twin or other
defect-generated boundaries may also be checked by varying the
preparation conditions.

Sometimes, instead of presenting the response of CDC materials to ac
electric fields via the dielectric constant, loss or conductivity,
the complex impedance is provided (e.g.,
\cite{Sinclair2002,Sebald2009}). This usually is done by showing
complex impedance-plane plots, i.e. $Z''(\nu)$ \textit{vs}.
$Z'(\nu)$. In such plots, any parallel RC circuit being part of the
equivalent circuit describing the sample (see, e.g., Figs.
\ref{fig:circuit} and \ref{fig:theo}(c)) will lead to a separate
semicircle (at least if their time constants $\tau=RC$ are well
separated). As this kind of plots allows for a simple graphical
evaluation (e.g., extrapolations of the semicircles enable the
determination of the involved resistors of the circuit) they were
often employed in earlier times when performing least-square fits of
complex functions were a difficult task. However, one has to be
aware that the frequency information is lost in this kind of plots
and nowadays it is state of the art to perform fits of the
dielectric spectra with the complete formula of the equivalent
circuit (e.g., Eqs. \ref{eq:Ct1} and \ref{eq:Ct2}), simultaneously
applied to complemental quantities as, e.g., dielectric constant and
conductivity or loss. Sometimes the observation of a second
semicircle in the impedance-plane plot, found in addition to the one
arising from the bulk, is taken as evidence for grain boundary or
other IBLC mechanisms. However, just in the same way as for an IBLC,
also an SBLC will lead to such a semicircle and no information
concerning the nature of the barrier layers is provided from such
type of plots.

Interfacial barrier effects usually lead to relaxation steps in
$\varepsilon'(T,\nu)$ with temperature-independent values of
$\varepsilon_{s}$. The static dielectric constant in case of MW
relaxations is determined by the interface capacitance $C_{i}$ (see
Sec. \ref{sec:interface}). The thickness of grain boundaries or of
depletion layers of diodes usually are not or only weakly
temperature dependent, thus explaining the constant
$\varepsilon_{s}(T)$. However, if the interface conductance $G_{i}$
should be of similar magnitude as the bulk value $G_{b}$, i.e. if
the condition $G_{i} \ll G_{b}$ mentioned in Sec.
\ref{sec:interface} is no longer fulfilled, the relation
$C'_{s}=C_{i}$ becomes invalid. In this case the low-frequency
dielectric constant $\varepsilon_{s}\propto C'_{s}$ may become
temperature dependent. Assuming $C_{i} \gg C_{b}$  and $G_{i}
\approx G_{b}$  we obtain from Eq. (\ref{eq:Cs}):

\begin{equation}
C'_{s}=\frac{G_{b}^{2}C_{i}}{(G_{i}+G_{b})^{2}}
\label{eq:Csrelaxor}
\end{equation}

\noindent The conductivity of bulk and of the barrier layer can be
assumed to exhibit strong semiconducting temperature dependence but
both quantities usually should not exhibit identical temperature
variation and in some temperature region the condition $G_{i} \ll
G_{b}$ may apply and in others $G_{i} \approx G_{b}$ might be valid.
Thus, a strong temperature dependence of the static dielectric
constant can result due to a transition from $C'_{s}=C_{i}$ to
$C'_{s}$ as given by Eq. \ref{eq:Csrelaxor}, which, e.g., for $G_{i}
= G_{b}$ would lead to $C'_{s}=C_{i}/4$. Under certain conditions,
this can lead to temperature characteristics of $\varepsilon_{s}$,
mimicking that of relaxor ferroelectrics.

\section{Colossal dielectric constants: Experimental results}
\label{sec:cdc}

In the following, we provide an overview of results from broadband
dielectric spectroscopy on various materials with CDCs as collected
in the Augsburg dielectric group during recent years, also including
new, so far unpublished data. For details on the dielectric
experiments and sample preparation, the reader is referred to our
earlier publications, e.g.,
\cite{Lunkenheimer2004,Krohns2008,Krohns2009a,Sebald2009,Krohns2009b,Schneider2001,Kant2008}.

\subsection{CaCu$_3$Ti$_4$O$_{12}$}
\label{sec:CCTO}

\begin{figure}[htb]
\centerline{\resizebox{.95\columnwidth}{!}{
\includegraphics{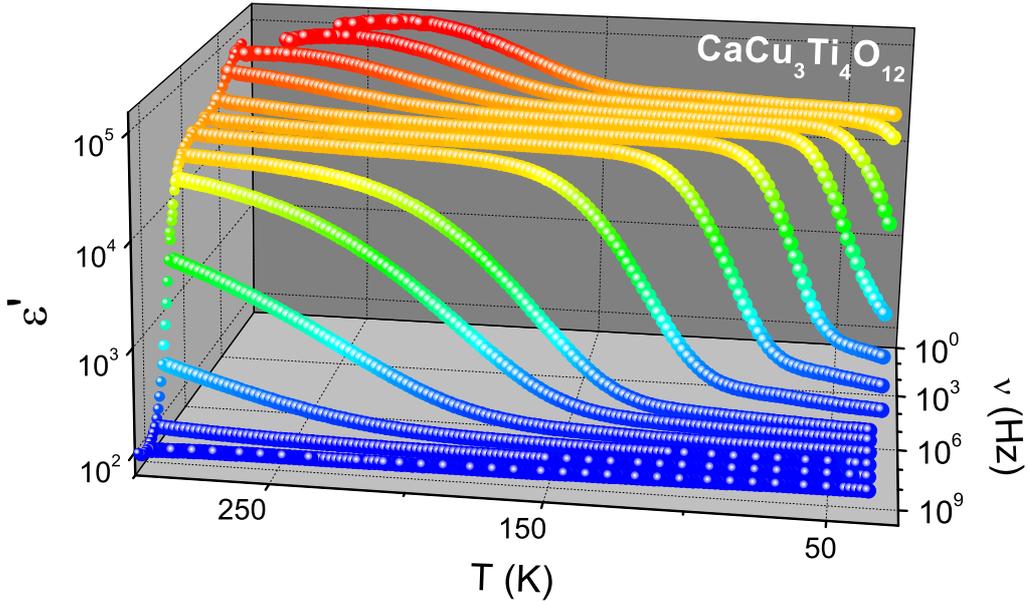} }}
\caption{Temperature- and frequency-dependent dielectric constant of
single-crystalline CCTO with silver-paint contacts.}
\label{fig:ccto3d}
\end{figure}

As mentioned above, CaCu$_3$Ti$_4$O$_{12}$ is the most prominent
material showing CDCs. The three-dimensional representation of Fig.
\ref{fig:ccto3d} provides a convenient overview of the temperature
and frequency dependence of its dielectric constant. Most
investigations on CCTO reported in literature were performed on
ceramic samples and there were also various efforts to prepare
high-quality thin films as a first step to application
\cite{Lin2002,Nigro2006,Fiorenza2009}. However, the measurements of
Fig. \ref{fig:ccto3d} were obtained on single-crystalline CCTO,
which is only rarely investigated so far
\cite{Homes2001,Krohns2007,Krohns2008}. As pointed out in the
original publications \cite{Subramanian2000,Homes2001,Ramirez2000},
in contrast to ferroelectric materials the CDC of CCTO remains
unchanged in a relatively broad temperature range. In Fig.
\ref{fig:ccto3d} this region corresponds to the plateau formed by
the yellow, orange and red data points. Indeed at the lowest
frequencies of the order of several Hz, this plateau extends over
the complete temperature range. However, at higher frequencies, this
range becomes successively restricted and at frequencies above some
100~MHz no CDCs are observed at all, even at room temperature
\cite{Lunkenheimer2004,Krohns2007}. The reason for this behaviour is
the strong relaxational mode typically observed in CCTO. It leads to
a step-like decrease of $\varepsilon'(T,\nu)$ with
\textit{de}creasing temperature or with \textit{in}creasing
frequency. Such characteristics is commonly found also in materials
with intrinsic relaxations due to dipolar degrees of freedom
\cite{Kremer2002}. For example, it closely resembles the findings in
glass forming liquids with dipolar molecules as, e.g., glycerol, the
only difference being the very large values of the static dielectric
constant reached at low frequencies and/or high temperatures
\cite{Lunkenheimer2000}.

\begin{figure}[htb]
\centerline{\resizebox{0.65\columnwidth}{!}{
  \includegraphics{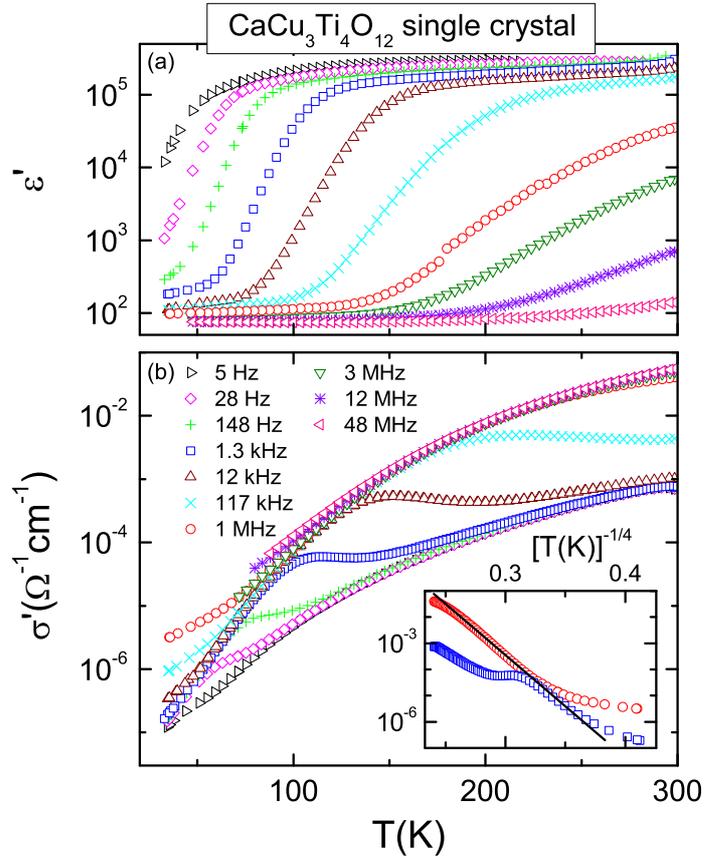} }}
\caption{Temperature-dependent dielectric constant (a) and
conductivity (b) of single-crystalline CCTO with sputtered gold
contacts at various frequencies \cite{Krohns2008}. The inset shows
the conductivity for 1.3~kHz and 1~MHz in VRH representation. The
solid line demonstrates VRH behaviour of the dc conductivity.}
\label{fig:CCTOTdep}
\end{figure}

In Fig. \ref{fig:CCTOTdep}(a) data on the same single crystal, but
now using sputtered contacts are provided in a conventional graph
showing the temperature dependence of $\varepsilon'$ for
measurements at different frequencies \cite{Krohns2008}. The
step-like decrease of $\varepsilon'(T)$ with low values at low
temperatures arises from a combination of the Debye
frequency-dependence shown in Fig. \ref{fig:debye} and the
semiconductor characteristics of the bulk conductance, which, via
$\tau=C_{i}/G_{b}$ (see Sec. \ref{sec:interface}), leads to a strong
increase of the relaxation time with decreasing temperature. Thus at
low temperatures, the system no longer can follow the ac field and
low $\varepsilon'$ values are observed. In Fig.
\ref{fig:CCTOTdep}(b) the corresponding conductivity curves are
shown. It should be noted that the conductivity is directly related
to the dielectric loss via
$\sigma'=\omega\varepsilon''\varepsilon_{0}$. Thus $\sigma'(T)$
curves are identical to those of $\varepsilon''(T)$, except for
their absolute values. However, the additional frequency factor
usually leads to a better distinguishability of the different curves
and thus to a better readability of the graphs. In addition, in
cases of typical interface-generated relaxational response the
conductivity representation provides further insight, e.g.,
concerning the behaviour of the intrinsic bulk conductivity. In
general, relaxation steps in $\varepsilon'$ should lead to
corresponding peaks in $\varepsilon''$ or $\sigma'$ (for
$\varepsilon''$, this is valid for both the temperature and the frequency
dependence, cf. Fig. \ref{fig:debye}). In Fig.
\ref{fig:CCTOTdep}(b) shoulders are observed instead of peaks
because of the additional contribution from the conductance of the
interface barrier, which is discussed in more detail below.

\begin{figure}[htb]
\centerline{\resizebox{0.65\columnwidth}{!}{
  \includegraphics{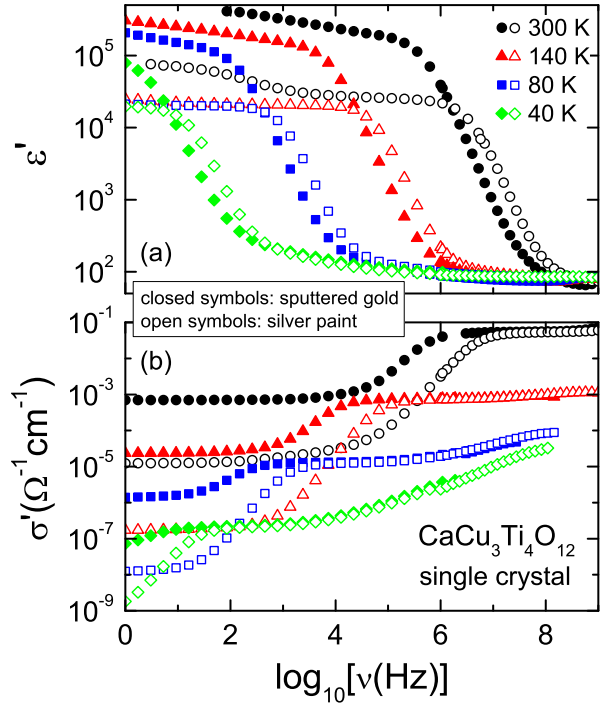} }}
\caption{Frequency-dependent dielectric constant (a) and
conductivity (b) of single-crystalline CCTO with silver-paint (open
symbols) and sputtered gold contacts (closed symbols) at selected
temperatures \cite{Krohns2008}.} \label{fig:CCTO_SC_AuAg}
\end{figure}

In Fig. \ref{fig:CCTO_SC_AuAg}, the frequency dependence of
dielectric constant and conductivity are shown for selected
temperatures. This figure contains the results from two different
measurements using silver paint or sputtered gold contacts applied
to the same sample. While at high frequencies the results from both
measurements agree, there are marked deviations at lower
frequencies. Especially, the static dielectric constant (Fig.
\ref{fig:CCTO_SC_AuAg}(a)) reaches values that are about one order
of magnitude larger for the sputtered contacts and the low-frequency
plateau of $\sigma'(\nu)$ (Fig. \ref{fig:CCTO_SC_AuAg}(b)) even
differs by about two decades. These curves are qualitatively similar
to the solid lines shown in the schematic plot of Fig.
\ref{fig:theo} and can be explained in the same way as discussed in
detail in Sec. \ref{sec:interface} in terms of interfacial effects.
Within this framework, at low frequencies the response is dominated
by the non-intrinsic barrier contribution. Obviously, for sputtered
contacts a much higher interfacial capacitance arises. As discussed
in detail in Sec. \ref{sec:distin}, this is just what is expected
for an SBLC effect caused by the formation of Schottky or MIS diodes
at the electrode-sample interfaces. Also the higher contact
conductance of the sputtered contacts, corresponding to the plateau
value at low frequencies in $\sigma'(\nu)$, can be understood in
this way. It should be noted that even in the sputtered case the
contact resistance, while lower than for silver paint, still is
considerable and clearly no ohmic contacts are formed. The reason
for the different contact conductances obtained for the two contact
materials is not the better quality of sputtered contacts. Instead
it is due to the variation in the formation of the diodes at the
electrodes, which is caused by the different "wetting" as discussed
in Sec. \ref{sec:distin}. Obviously, the much higher resistance of
the air gaps, replacing parts of the diode depletion layers in the
silver-paint case, must play a role here.

According to the equivalent-circuit framework developed in Secs.
\ref{sec:interface} and \ref{sec:distin}, at high frequencies the
bulk response should be detected, which is nicely corroborated by
the agreement of the curves from both measurements shown in Fig.
\ref{fig:CCTO_SC_AuAg}. At 40~K and 80~K, in Fig.
\ref{fig:CCTO_SC_AuAg}(b) the bulk conductivity increases with
frequency, approaching a power law for the highest frequencies. As
discussed in Sec. \ref{sec:hop}, this is the typical UDR arising
from hopping of localised charge carriers
\cite{Long1982,Elliott1987,Jonscher1983,Mott1979}. The corresponding
contribution in $\varepsilon'(\nu)$ is the slight increase with
decreasing frequency, best seen in the 40~K curve between about
100~Hz and 100~kHz (cf. dashed line in Fig. \ref{fig:theo}(a)).
Coming now back to Fig. \ref{fig:CCTOTdep}(b), the merging of the
conductivity curves for different frequencies at low and high
temperatures can be understood as follows: The lower merging curve
corresponds to the low-frequency plateau seen in the frequency
dependence (Fig. \ref{fig:CCTO_SC_AuAg}(b)), i.e. it mirrors the
temperature dependence of the contact conductance. The approach of
this curve with increasing temperature prevents the observation of
well-defined peaks in $\sigma'(T)$, which would become visible only
for much smaller contact conductance (see, e.g., Fig. 1 in Ref.
\cite{Krohns2008}). The upper merging curve in Fig.
\ref{fig:CCTOTdep}(b) corresponds to the intrinsic conductivity of
CCTO. At $T<75$~K deviations arise for the highest frequencies,
which signifies just the same UDR contributions as revealed in Fig.
\ref{fig:CCTO_SC_AuAg}(b). The intrinsic dc conductivity of CCTO
does not follow conventional thermally activated behaviour. Instead
it can be quite well described by the prediction of the VRH model
\cite{Mott1979}, i.e. $\sigma_{dc}\propto T^{1/4}$
\cite{Krohns2008,Zhang2004,Tselev2004}. This is demonstrated in the
inset of Fig. \ref{fig:CCTOTdep} where the x-axis was chosen to
linearise the $T^{1/4}$ law of the VRH model. The curves for 1~MHz
and 1.3~kHz, shown in the inset, correspond to the intrinsic dc
conductivity for high and low temperatures, respectively. As
demonstrated by the solid line, indeed VRH provides a good fit of
the data. This is fully consistent with the detection of the typical
UDR power law in the frequency dependence, which is characteristic
for hopping transport.

\begin{figure}[htb]
\centerline{\resizebox{0.65\columnwidth}{!}{
  \includegraphics{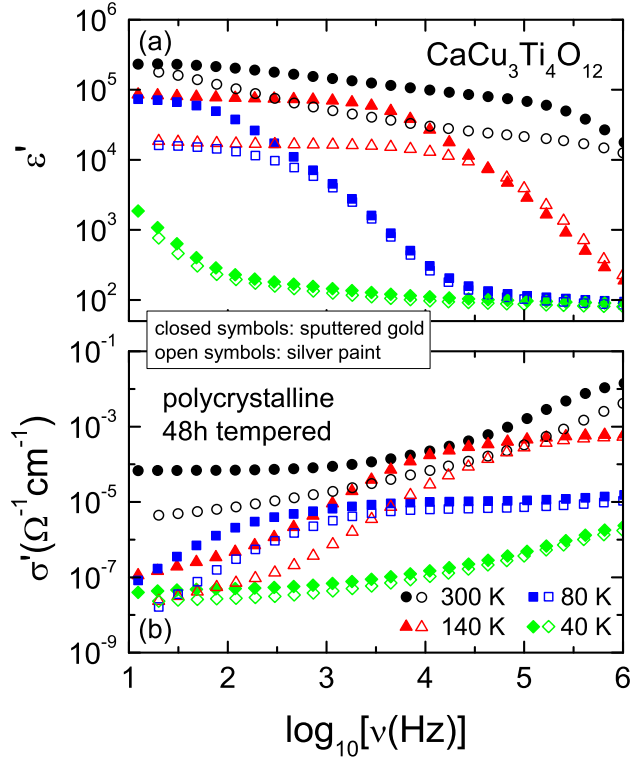} }}
\caption{Frequency-dependent dielectric constant (a) and
conductivity (b) of polycrystalline CCTO (48~h tempered) with
silver-paint (open symbols) and sputtered gold contacts (closed
symbols) at selected temperatures.} \label{fig:CCTO_PC_AuAg}
\end{figure}

The strong dependence of the CDC of CCTO on contact preparation is
also found in polycrystals as demonstrated in Fig.
\ref{fig:CCTO_PC_AuAg}. Again, strong deviations show up at low
frequencies while at high frequencies the bulk response governs the
spectra for both types of contact. Taking together all these
results, it seems clear that the CDCs in CCTO to a large extent are
determined by surface-related effects. However, it should be noted
that experiments with varying contacts were reported in literature
that partly do not allow for such definite conclusions
\cite{Adams2006,Yang2005,Ferrarelli2009}. Surface variations
introduced by annealing, polishing and other surface treatments
obviously also can play an important role, in addition to the effect
of the contact material, making the interpretation of experimental
results difficult. Finally, it should be noted that for the ceramic
sample with sputtered contacts, CDC values of the order of $10^{5}$
are observed (Fig. \ref{fig:CCTO_PC_AuAg}(a)). Usually such high
magnitudes of $\varepsilon'$ are found for CCTO single crystals only
and for polycrystalline samples often much lower values are revealed
\cite{Subramanian2000,Ramirez2000,Lunkenheimer2004}. However,
annealing the sample for 48h to achieve large grains (see below) and
using sputtered contacts to reach good contact wetting obviously can
lift the dielectric constant of ceramic samples into similar regions
as for single crystals.

\begin{figure}[htb]
\centerline{\resizebox{0.65\columnwidth}{!}{
  \includegraphics{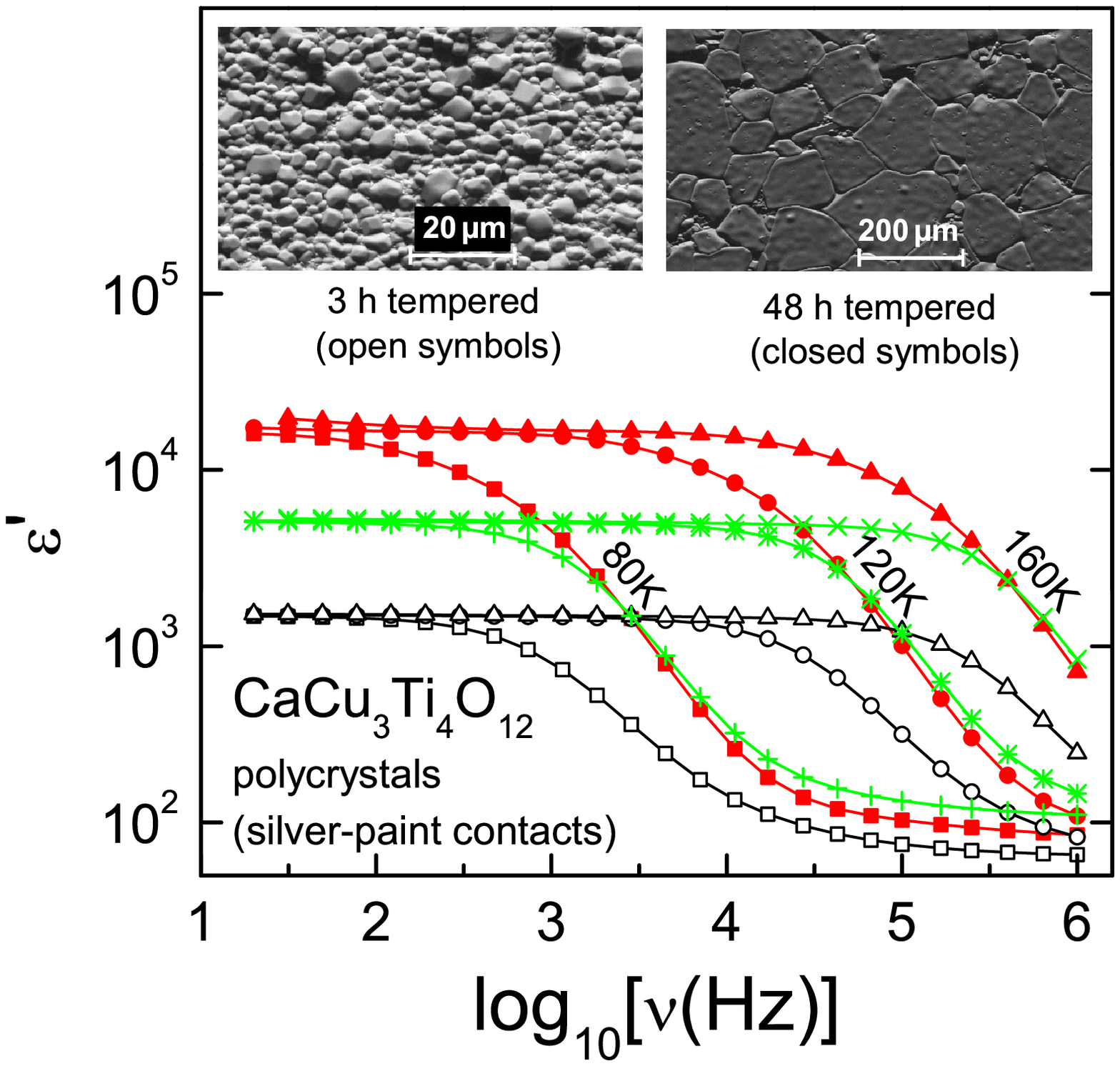} }}
\caption{Frequency-dependent dielectric constant of ceramic CCTO
samples tempered for 3 h (open symbols), 24 h (symbols) and 48 h
(closed symbols) with silver-paint contacts \cite{Krohns2008}. The
insets show the surface topographies obtained by ESEM of the 3 h and
48 h tempered samples.} \label{fig:CCTO_tempered}
\end{figure}

As mentioned in Sec. \ref{sec:distin}, sintering of ceramic samples
to achieve different grain sizes is a quite commonly employed way of
checking for grain-boundary related IBLC effects (e.g.,
\cite{Adams2002,Bender2005,Zang2005}). Figure
\ref{fig:CCTO_tempered} shows the frequency-dependent dielectric
constant of ceramic CCTO that has been subjected to tempering in air
at 1000~$^\circ$C for 3, 24 and 48 hours \cite{Krohns2008}. With
increasing tempering time, i.e. increasing grain sizes, the
magnitude of the CDC increases continuously, while the intrinsic
value of $\varepsilon'$, read off at high frequencies, remains
nearly unaltered. At first glance, this finding seems to strongly
point to grain boundaries as the origin of the CDCs as the overall
thickness of the interfaces between grains becomes less in relation
to the sample thickness and thus the corresponding capacitance
increases (see Sec. \ref{sec:distin}). However, as mentioned in Sec.
\ref{sec:distin} also the surface smoothness may change with
sintering and therefore an SBLC effect cannot be fully excluded
\cite{Krohns2008}. The effect of sintering on the sample surface is
seen in the inset of Fig. \ref{fig:CCTO_tempered} providing surface
topographies detected by Environmental Scanning Electron Microscopy
(ESEM). Indeed for the sample tempered for 48~h, the larger grains
also lead to a much smoother surface. In any case, there are several
sophisticated experiments reported in literature that seem to
suggest that grain boundary or other SBLC mechanisms at least play
some role for the generation of the CDCs
\cite{Chung2004,Adams2006,Bender2005,Fiorenza2008,Ferrarelli2009}. A
solution of the partly contradictory results in literature could be
a second relaxation, which was reported in several publications to
occur in ceramic samples
\cite{Lunkenheimer2004,Zhang2005,Wang2006,Krohns2007,Krohns2008} (cf. dotted
line in Fig. \ref{fig:theo}). Thus, one may assume that one
relaxation is due to SBLCs and the second one is caused by IBLCs.
However, in Ref. \cite{Krohns2007} it was shown that the second
relaxation also shows up in single crystals, excluding a grain
boundary mechanism. In fact, this relaxation is also weakly seen in
Fig. \ref{fig:CCTO_SC_AuAg}(a) as an additional step-like increase
below about 10~kHz revealed by the curve measured at 300~K with
silver-paint contacts. It also may well be possible that the
domination of the dielectric response by IBLC or SBLC effects is
strongly sample dependent and no clear-cut statement for CCTO in
general can be made. For example, in Ref. \cite{Ferrarelli2009} the
prevailance of SBLC effects for single crystals and of grain
boundary effects for ceramic samples was proposed. Differences may
also arise for fine and coarse-grained ceramics or for varying
surface resistivities \cite{Yang2005}.

\begin{figure}[htb]
\centerline{\resizebox{0.75\columnwidth}{!}{
  \includegraphics{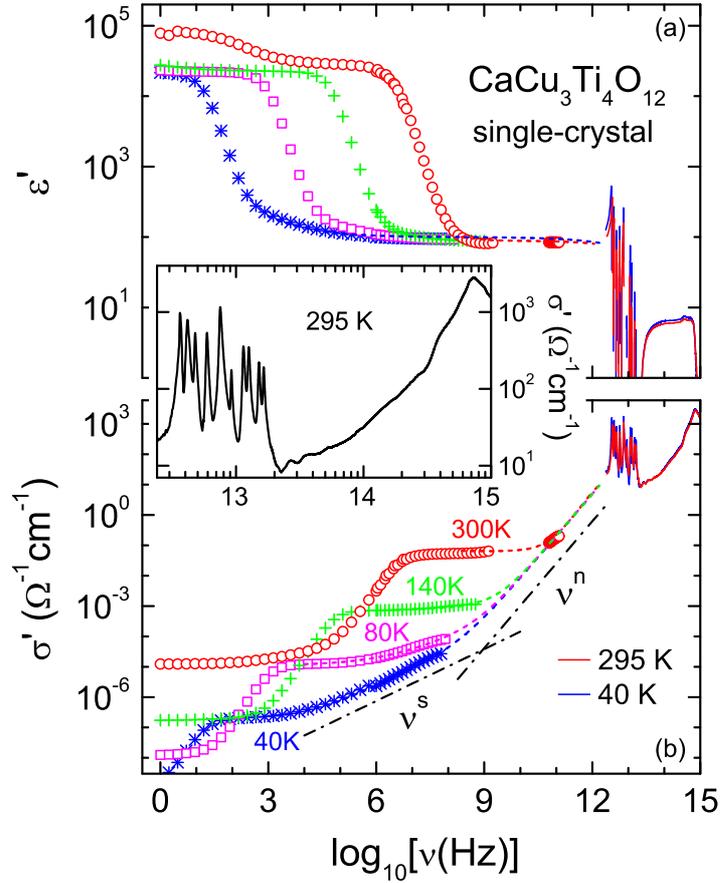} }}
\caption{Dielectric constant (a) and dynamic conductivity (b) of
CCTO over 15 decades in frequency for various temperatures
\cite{Kant2008}. The symbols at $\nu < 10^{12}$~Hz and the lines at
$\nu > 10^{12}$~Hz show experimental data. Dashed lines are fits of
the data beyond the relaxation, taking into account dc and ac
conductivities, the latter described by sub- and superlinear power
laws, $\nu^s$ and $\nu^n$. The inset shows the dynamic conductivity
of CCTO at 295 K in the optical regime.} \label{fig:CCTO_optic}
\end{figure}

As revealed by Figs. \ref{fig:CCTOTdep} - \ref{fig:CCTO_tempered},
the intrinsic dielectric constant of CCTO, which can be read off at
high frequencies and/or low temperatures is of the order of 100 and
thus relatively high. HfO$_2$, the most considered
high-$\varepsilon'$ material for applications in integrated
circuits, e.g., has $\varepsilon\approx 22$ only. For
heterogeneity-dominated systems, high bulk $\varepsilon'$-values may
enhance the CDCs. For example in case of diode-generated SBLCs the
capacitance of the depletion layer, which determines the magnitude
of the CDC of course depends on the dielectric constant in this
region. The depletion layer of course still is CCTO, only depleted
from any mobile charge carriers. Its $\varepsilon'$, determined by
the ionic and electronic polarisability thus should be of similar
magnitude at the intrinsic $\varepsilon'$ of CCTO. Therefore the CDC
should be directly proportional to the bulk value.  To understand
the generation of the high bulk $\varepsilon'$ in CCTO, Fig.
\ref{fig:CCTO_optic} shows combined spectra of dielectric, THz and
infrared measurements \cite{Kant2008}. As revealed by Fig.
\ref{fig:CCTO_optic}(a), in the narrow frequency region between
about 3 and 20~THz, $\varepsilon'(\nu)$ decreases by about one order
of magnitude. In this region a number of sharp resonances show up,
caused by phononic excitations. Thus, one can conclude that the high
intrinsic dielectric constant of CCTO mainly arises from the ionic
polarisability. This notion is corroborated by a quantitative
evaluation of the phonon modes performed by fitting the spectra
using ten oscillators \cite{Kant2008}. The resulting sum of the
dielectric strengths of all phonon modes is about 77 at room
temperature and about 96 at 5~K; the dielectric constant due to
electronic polarisability is 6.5. Thus we obtain for the static
dielectric constant of the infrared experiment at room temperature
$\varepsilon_{s,o}\approx84$, which agrees well with the intrinsic
dielectric constant of the dielectric experiments,
$\varepsilon_{\infty}\approx83$. With $\varepsilon'\approx85$ also
the result from THz spectroscopy matches these values.

In the combined plot of the conductivity of Fig.
\ref{fig:CCTO_optic}(b), the sublinear UDR power law arising from
hopping conductivity is nicely seen (cf. lower dash-dotted line) but
also additional contributions are detected. When also taking into
account the THz and infrared results, another steeper power law
$\sigma'\propto \nu^{n}$ with exponent $n\approx 1.4$ shows up (cf.
upper dash-dotted line). Indeed, fits assuming the sum of dc
conductivity, a sublinear and a superlinear power law  (dashed
lines) are able to describe the spectra beyond the relaxation mode.
It should be noted that via the Kramers-Kronig relation, a
superlinear power law in $\sigma'(\nu)$ of course also leads to a
contribution in $\varepsilon'(\nu)$, namely an additional decrease
proportional to -$\nu^{n-1}$. This decrease is too small to actually
show up within the resolution of the experimental data. However, it
is observed in the fits shown by the dashed lines in Fig.
\ref{fig:CCTO_optic}(a) as a slight decrease between 1~GHz and
1~THz. A superlinear power law, showing up at frequencies beyond the
validity of the sublinear UDR power law, was previously observed in
various transition-metal oxides and other materials
\cite{Lunkenheimer2003,Lunkenheimer2003b}. It is clear that
additional contributions beyond the UDR must prevail in the region
bridging the gap between dielectric spectroscopy
($\nu\lesssim1$~GHz) and infrared experiments ($\nu\gtrsim1$~THz)
because in most materials a simple extrapolation of the UDR does not
match the absolute values of $\sigma'$ at the lowest frequencies of
the infrared experiments. In Ref. \cite{Lunkenheimer2003b} the
superlinear $\nu^{n}$ power law was proposed to be a universal
property of disordered matter but its microscopic origin still is
unclarified. Interestingly, for high frequencies a transition from
phonon-assisted to photon-assisted hopping with
$\sigma'\propto\nu^{2}$ is predicted for a Fermi glass
\cite{Mott1979,Shklovskii1981}. However, this transition should
appear at $h\nu>k_{B}T$, which is not fulfilled here.

The inset of Fig. \ref{fig:CCTO_optic} shows a magnified view of the
results in the infrared region. At low frequencies, $\nu<20$~THz,
ten phonon modes are clearly revealed. As discussed in detail in
Ref. \cite{Kant2008}, the low-lying modes show a decrease of their
eigenfrequencies with decreasing temperature, i.e. their temperature
dependence resembles the soft-phonon behaviour that is typical for
ferroelectric materials \cite{Lines1979}. Thus, while CCTO clearly
is not ferroelectric, at least some ferroelectric correlations seem
possible, which also is consistent with the increase of the bulk
dielectric constant from 85 at room temperature to about 100 at 5~K
revealed by both the dielectric and the infrared experiments
\cite{Kant2008}. Beyond 20~THz, the further increase of
$\sigma'(\nu)$ is due to electronic excitations. They qualitatively
agree with the predictions from LSDA band structure calculations
\cite{He2002} assuming transitions from filled hybridised O~2$p$ and
Cu~3$d$ bands to empty O~2$p$/Cu~3$d$ and Ti~3$d$ states
\cite{Kant2008}.

\subsection{CaCu$_3$Ti$_4$O$_{12}$-related systems}
\label{sec:CCTOrel}

\begin{figure}[htb]
\centerline{\resizebox{0.65\columnwidth}{!}{
  \includegraphics{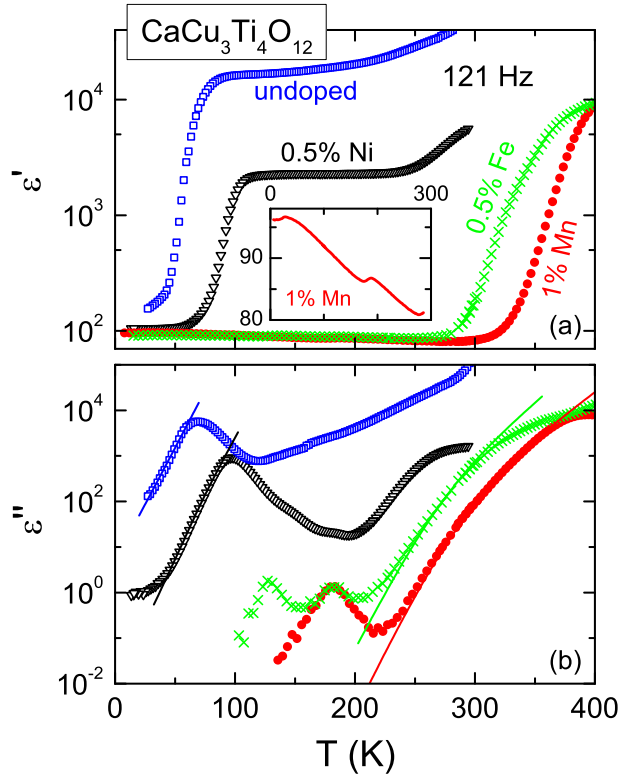} }}
\caption{Temperature-dependent dielectric constant (a) and
dielectric loss (b) of undoped and doped (1\%~Mn, 0.5\%~Fe and
0.5\%~Ni) polycrystalline CCTO at 121 Hz \cite{Krohns2009a}. The
lines in (b) indicate the temperature-dependent development of the
dc conductivity contribution, detected at the low-frequency flank of
the MW relaxation peaks. The inset shows a magnified view of
$\varepsilon'(\nu)$ for 1\% Mn doping below room temperature.}
\label{fig:CCTO_doped}
\end{figure}

To make use of the CDCs of CCTO for technical applications but
avoiding the disadvantages of this materials as, e.g., its
relatively high dielectric losses, one can proceed along different
routes. Aside of an optimisation of the pure material, for example
by adjusting its microstructure, doping of CCTO or the complete
substitution with different elements may seem a promising approach.
In Fig. \ref{fig:CCTO_doped} we compare the complex permittivity at
121~Hz of pure and three doped CCTO samples \cite{Krohns2009a}. It
was previously reported that doping CCTO with Fe or Mn leads to a
marked variation of its dielectric constant and a reduction of its
dc conductivity
\cite{Kobayashi2003,Grubbs2005,Chung2006,Li2006,Li2007,Brize2009}.
In Fig. \ref{fig:CCTO_doped}(a) the typical MW relaxation step is
seen for all investigated materials with CDC values varying between
2000 and 17000. However, for the iron- and manganese-doped samples,
the relaxation step is shifted to much higher temperatures than for
the undoped and nickel-doped samples. Quite in general, shifts of
relaxational features to higher temperatures correspond to shifts to
lower frequencies in frequency-dependent plots and thus (because
$\tau\propto 1/\nu_p$ with $\nu_p$ the loss-peak frequency) imply an
increase of the relaxation times. Obviously, for the Mn- and
Fe-doped samples, already shortly below 400~K the relaxation time
becomes too large for the systems to follow the excitation frequency
of 121 Hz. In contrast, for the pure and Ni-doped sample the
relaxation time is much smaller and the materials can follow the
frequency for temperatures as low as 100~K before
$\varepsilon'(\nu)$ finally relaxes to the intrinsic bulk value of
the order of 100.

In Fig. \ref{fig:CCTO_doped}(b) the loss peaks corresponding to the
MW relaxation are nicely revealed for the undoped and Ni-doped
sample. For the Fe- and Mn-doped sample this is not the case and the
MW relaxation peak shows up as shoulder at 300-400~K only. There the
situation is similar as, e.g., seen in Fig. \ref{fig:CCTOTdep}(b)
for pure CCTO, which is discussed in detail in Sec. \ref{sec:CCTO}.
As noted above, the marked shift of the MW-relaxation features
implies a strong variation of the relaxation time. As
$\tau=C_{i}/G_{b}$ (see Sec. \ref{sec:interface}), and a variation
of $C_{i}$ is unlikely, the conductivity must be strongly reduced
for Fe and Mn doping. This is well corroborated by Fig.
\ref{fig:CCTO_doped}(b) where the left flanks of the MW-relaxation
peaks (indicated by the solid lines) are directly proportional to
the dc conductivity (see Sec. \ref{sec:interface}). Obviously,
already for relatively small doping levels the dc conductivity
becomes reduced by many orders of magnitude when doping CCTO with
iron or manganese. Most likely, charge transport in pure CCTO mainly
arises from slight oxygen deficiencies, which are compensated by
doping with manganese or iron, which should be substituting on the
titanium site \cite{Krohns2009a}. In contrast, Ni doping, most
likely of isoelectronic type substituted on the copper-place
\cite{Krohns2009a}, leads to a much smaller reduction of
conductivity and thus a smaller shift of the MW relaxation only.
These findings demonstrate that reducing the intrinsic bulk
conductivity of a material to minimise its dielectric losses is not
a suitable approach for its optimisation for technical application
in case of a MW-generated mechanism. Reducing the bulk conductance
will lead to an increase of the relaxation time and, thus, to a
strong restriction of the frequency range available for application.
For example, for 1\% Mn doping, even at a frequency as low as 121~Hz
no CDCs are found anymore at room temperature (Fig.
\ref{fig:CCTO_doped}(a)). Instead, reducing the conductance of the
barrier layers seems to be the more feasible way of optimising
MW-dominated CDC materials. This should reduce the losses without
affecting the relaxation time. However, a simple increase of the
layer thickness will not help as this would lead to a simultaneous
reduction of the CDC values.

For the iron- and manganese-doped samples, in Fig.
\ref{fig:CCTO_doped}(b) two, respectively one additional relaxation
peak is revealed at temperatures below the dominating MW loss-peak.
Interestingly, the peak located at about 180~K is found for both
materials and an evaluation of the frequency-dependent loss reveals
that the relaxation time associated with this feature shows
identical temperature dependence for both materials
\cite{Krohns2009a}. The inset of Fig. \ref{fig:CCTO_doped}
demonstrates for the Mn-doped sample that the intrinsic relaxation
also shows up in the real part as a small step superimposed to the
general decreasing trend of $\varepsilon'(\nu)$.
 A closer look at the curve for the Ni-doped sample in Fig.
\ref{fig:CCTO_doped}(b) may also reveal a similar feature for this
material. Thus one can suspect that this relaxation is an intrinsic
property of CCTO and not related to the doping. It should be noted
that for pure CCTO, the intrinsic dielectric properties in this
temperature region are inaccessible due to the dominating
non-intrinsic barrier contribution (cf. Fig. \ref{fig:CCTOTdep}),
except for very high frequencies when the MW relaxation is shifted
towards high temperatures. However, in this case, also the intrinsic
relaxation should be shifted and still may remain undetected. Only
the low bulk conductivity of the doped compounds and the subsequent
shift of the relaxation feature to higher frequencies could reveal
this intrinsic relaxation. Its microscopic origin is unclear at
present. The second relaxation in the Fe-doped sample does not show
up for Mn doping and thus seems to be related to the Fe defects
\cite{Grubbs2005}.

As revealed by the inset of Fig. \ref{fig:CCTO_doped}, aside of the
small relaxation step, the bulk $\varepsilon'(T)$ shows a clear
increase with decreasing temperature. This agrees with the findings
from measurements at GHz in pure CCTO where the frequency was
sufficiently high to shift the MW-relaxation step to temperatures
beyond room temperature and also with the results from an analysis
of the phonon modes measured by infrared spectroscopy
\cite{Kant2008}. As discussed above, this finding may indicate
ferroelectric-like correlations in CCTO and there are even
speculations about incipient ferroelectricity in doped CCTO
\cite{Li2007}, similar to SrTiO$_{3}$ \cite{Viana1994}. Another
interesting finding is the peak of $\varepsilon'(T)$ occurring just
at the transition into an antiferromagnetic state below about 25~K.
This small but significant magnetocapacitive effect is discussed in
detail in Ref. \cite{Krohns2009a}.

Finally, we want to mention the additional increase of
$\varepsilon'(T)$ observed at high temperatures, $T\gtrsim200$~K, in
the pure and Ni-doped sample. It signifies the presence of a second
relaxation, in addition to the main MW relaxation, which also is of
non-intrinsic origin, as briefly discussed above. For a detailed
treatment of the second relaxation, see Ref. \cite{Krohns2008}.

\begin{figure}[htb]
\centerline{\resizebox{0.55\columnwidth}{!}{
  \includegraphics{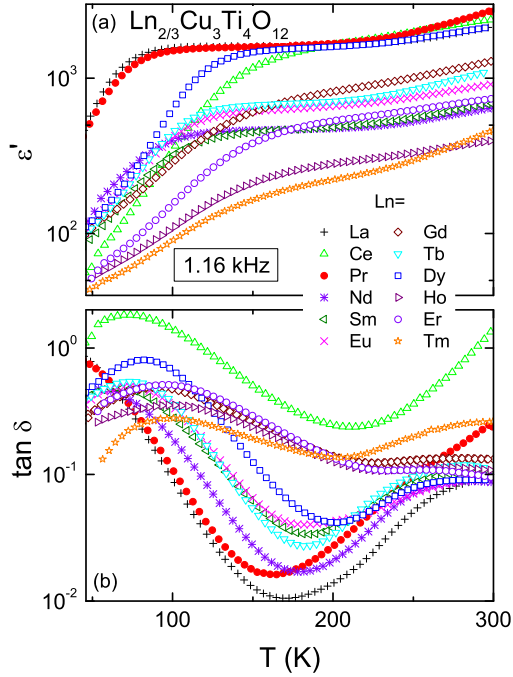} }}
\caption{Temperature-dependent dielectric constant (a)
\cite{Sebald2009} and loss tangent (b) of the investigated
LnCTO-compounds at 1.16 kHz (silver-paint contacts).}
\label{fig:LnCTO}
\end{figure}

\begin{figure}[htb]
\centerline{\resizebox{0.55\columnwidth}{!}{
  \includegraphics{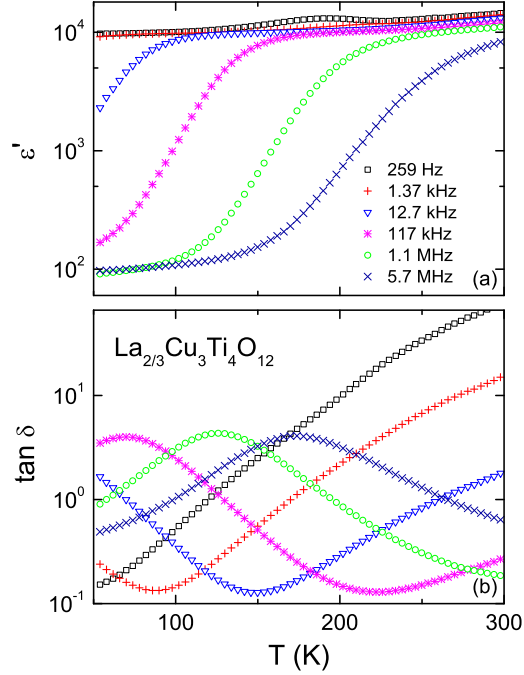} }}
\caption{Temperature-dependent dielectric constant (a) and loss
tangent (b) of ceramic La$_{2/3}$Cu$_{3}$Ti$_{4}$O$_{12}$ with
silver-paint contacts for various frequencies.} \label{fig:LaCTO}
\end{figure}

Already in the pioneering work on CCTO by Subramanian \textit{et
al}. and a follow-up paper \cite{Subramanian2000,Subramanian2002}, a
number of compounds, isostructural to CCTO were introduced. Until
now, only few of them are satisfactorily characterised by dielectric
spectroscopy and it is not clear why CCTO should be unique within
this large group of materials. Indeed, dielectric constants of
colossal magnitude were reported for some of these compounds
\cite{Subramanian2002,Sebald2009,Parkash2006,Parkash2008,Liu2004,Liu2005,Zhou2005,Ferrarelli2006}.
Figure \ref{fig:LnCTO} shows the dielectric constant (a) and the
loss tangent ($=\varepsilon''/\varepsilon'$) at 1.16~kHz (b) for a
number of ceramic Ln$_{2/3}$Cu$_{3}$Ti$_{4}$O$_{12}$ samples where
Ca was replaced by various lanthanides \cite{Sebald2009}. In all
cases, the typical interface-barrier induced MW relaxation is
observed with a high-temperature plateau in $\varepsilon'(T)$ of
large magnitude, $\varepsilon'_{s}$ reaching colossal values
exceeding 1000 for Ln~=~La, Ce, Pr and Dy. Among the latter, the
cerium compound has the highest loss tangent and seems less suited
for application. The relaxation features of the dysprosium compound
arise at clearly higher temperatures than for Ln~=~La and Pr. Thus
its relaxation time is higher and the CDC can be expected to occur
in a smaller frequency range only (cf. the above discussion of the
shift of the relaxations in Fig. \ref{fig:CCTO_doped}).
La$_{2/3}$Cu$_{3}$Ti$_{4}$O$_{12}$ and
Pr$_{2/3}$Cu$_{3}$Ti$_{4}$O$_{12}$, which behave nearly identical in
Fig. \ref{fig:LnCTO}, seem the most promising materials and in Fig.
\ref{fig:LaCTO} we provide a more detailed plot of the dielectric
properties of the lanthanum compound. For an in-depth treatment of
the dielectric properties of Pr$_{2/3}$Cu$_{3}$Ti$_{4}$O$_{12}$, see
Ref. \cite{Sebald2009}.

Figure \ref{fig:LaCTO} shows the temperature dependence of the
dielectric constant and loss tangent of
La$_{2/3}$Cu$_{3}$Ti$_{4}$O$_{12}$ for a number of frequencies.
Dielectric properties of this material were also reported in Refs.
\cite{Parkash2006,Parkash2008,Zhou2005}. The overall behaviour of
$\varepsilon'(T)$ (Fig. \ref{fig:LaCTO}(a)) closely resembles that
of pure CCTO as shown, e.g., in Fig. \ref{fig:CCTOTdep}. In contrast
to the measurements of La$_{2/3}$Cu$_{3}$Ti$_{4}$O$_{12}$, included
in Fig. \ref{fig:LnCTO}, a different sample was used, which was
treated to increase the grain size. In this sample, indeed a truly
colossal magnitude of $\varepsilon'$ of about 10000 is achieved.
Interestingly, it seems to arise from an IBLC effect as this value
remained nearly unaffected by the type of electrodes used in the
measurements (just as for Pr$_{2/3}$Cu$_{3}$Ti$_{4}$O$_{12}$; see
Ref. \cite{Sebald2009} for details). The intrinsic $\varepsilon'$,
read off at low temperatures and high frequencies is about 100,
similar as in CCTO. The loss tangent exhibits peaks, which, just as
for $\varepsilon''(T)$, are characteristic features of relaxations
(see Sec. \ref{sec:interface}). The additional increase at high
temperatures arises from the conductance of the barriers, just as
the lower merging curve revealed in the conductivity plot of pure
CCTO (Fig. \ref{fig:CCTOTdep}(b); see also discussion of this figure
in Sec. \ref{sec:CCTO}). Overall, here we have a material that has
properties at least as good as CCTO. A similar statement can be made
for Pr$_{2/3}$Cu$_{3}$Ti$_{4}$O$_{12}$ \cite{Sebald2009}. Therefore
it seems that there is nothing peculiar in CCTO and there may be
many more isostructural materials with comparable or even better
dielectric properties.

Another promising member of the family of CCTO-related materials is
Cu$_{2}$Ta$_{4}$O$_{12}$. Its crystal structure is derived from that
of CCTO by leaving the Ca sites unoccupied and replacing 1/3 of the
copper sites by vacancies. Its dielectric behaviour was reported to
be similar to that of CCTO, reaching CDCs of about 75000
\cite{Renner2004}. For details the reader is referred to Ref.
\cite{Renner2004}.

\subsection{La$_{2-x}$Sr$_{x}$NiO$_4$}
\label{sec:LSNO}

\begin{figure}[htb]
\centerline{\resizebox{0.65\columnwidth}{!}{
  \includegraphics{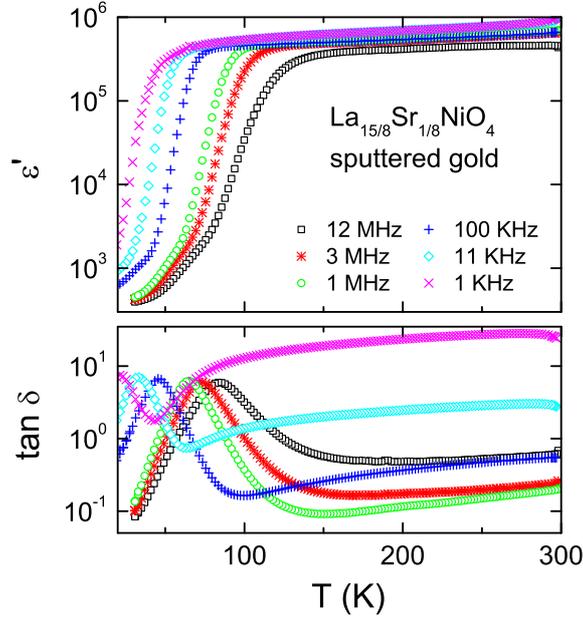} }}
\caption{Temperature-dependent dielectric constant (a) and loss
tangent (b) of a LSNO single crystal with sputtered gold contacts
for various frequencies \cite{Krohns2009b}.} \label{fig:LSNO_Tdep}
\end{figure}

As mentioned in Sec. \ref{sec:interface}, spontaneously arising
interfaces caused, e.g., by electronic phase separation or charge
order are of high interest for possible applications of CDC
materials for capacitive circuit elements. Interestingly there are
some reports on CDCs in the system La$_{2-x}$Sr$_{x}$NiO$_{4}$
\cite{Rivas2004,Park2005,Liu2008}, which is known to exhibit
electronic phase separation, namely a stripe-like ordering of holes,
in large portions of its phase diagram
\cite{Chen1993,Sachan1995,Lee1997,Tranquada1996,Yamanouchi1999,Du2000,Lee2002,Tranquada2002,Ishizaka2004}.
In Fig. \ref{fig:LSNO_Tdep} the temperature dependences of the
dielectric constant \cite{Krohns2009b} and the loss tangent of
single-crystalline La$_{15/8}$Sr$_{1/8}$NiO$_{4}$ (LSNO) are
provided. Typical relaxation steps (a) and peaks (b), just as in
CCTO, are revealed with extremely large values of
$\varepsilon_{s}\approx 600000$. The loss tangent at high
frequencies is comparable to or even lower than that of CCTO
\cite{Homes2001}.

\begin{figure}[htb]
\centerline{\resizebox{0.65\columnwidth}{!}{
  \includegraphics{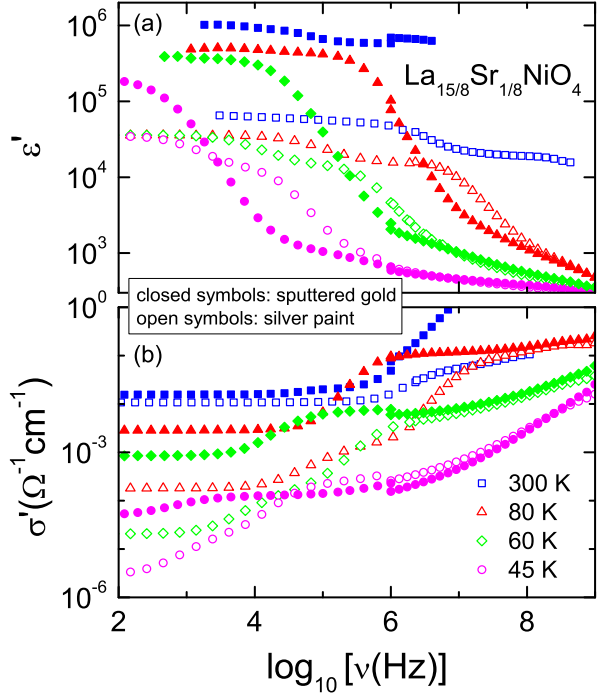} }}
\caption{Frequency-dependent dielectric constant (a) and
conductivity (b) of single-crystalline LSNO with silver-paint
(open symbols) and sputtered gold contacts (closed symbols) at
selected temperatures \cite{Krohns2009b}.} \label{fig:LSNO_AuAg}
\end{figure}

To elucidate the possible role of SBLCs for the generation of the
observed CDCs, in Fig. \ref{fig:LSNO_AuAg} the frequency dependence
of the dielectric constant (a) and the conductivity (b) are provided
for both, contacts prepared by sputtering (as in Fig.
\ref{fig:LSNO_Tdep}) and by applying silver paint. Just as for CCTO
single- and polycrystals (cf. Figs. \ref{fig:CCTO_SC_AuAg} and
\ref{fig:CCTO_PC_AuAg}), a marked variation of $\varepsilon_{s}$ is
found with much higher values for the sputtered contacts. Thus, it
is clear that surface generated MW effects indeed do contribute to
the CDCs in LSNO. Again, at high frequencies the curves obtained
with different contact types agree, showing the intrinsic response
(for 300~K this regime is reached beyond the investigated frequency
range only). The intrinsic $\sigma'(\nu)$ (Fig.
\ref{fig:LSNO_AuAg}(b)) is governed by a contribution from dc
conductivity (e.g., the approximate plateau at about
$2\times10^{-4}~\mathrm{\Omega}^{-1}\mathrm{cm}^{-1}$ in the 45~K
curves) and a marked UDR power-law increase due to hopping charge
transport (see Sec. \ref{sec:hop}). In $\varepsilon'(\nu)$ (Fig.
\ref{fig:LSNO_AuAg}(a)), in contrast to CCTO no clear saturation at
high frequencies is seen. Obviously, up to the highest frequency the
UDR contribution, $\varepsilon'\propto\nu^{s-1}$, is larger than
$\varepsilon_{\infty}$ arising from the ionic and electronic
polarisabilites. This leads to relatively high values of the
intrinsic bulk values of $\varepsilon'$ at high frequencies, e.g.,
$\varepsilon'(1~\mathrm{GHz})\approx300$.

The $\varepsilon'$ spectra obtained with silver-paint contacts shown
in Fig. \ref{fig:LSNO_AuAg} reveal clear indications for a second
relaxation step at low frequencies, quite similar to CCTO
\cite{Krohns2007}. Thus, a further mechanism enhancing
$\varepsilon'$, in addition to the suspected SBLCs, must be active
in this material. In Refs. \cite{Krohns2009b,Krohns2010} it was
speculated that only the second, low-frequency relaxation may be due
to an SBLC effect. In the spectra obtained for sputtered contacts,
this second relaxation dominates the response and the first,
high-frequency relaxation may become undetectable. Then a
contribution from the electronic phase separation may well be
possible, leading to a distinct separate relaxation step for the
silver-paint sample only . Clearly further work is necessary to
corroborate this scenario.

\begin{figure}[htb]
\centerline{\resizebox{0.8\columnwidth}{!}{
  \includegraphics{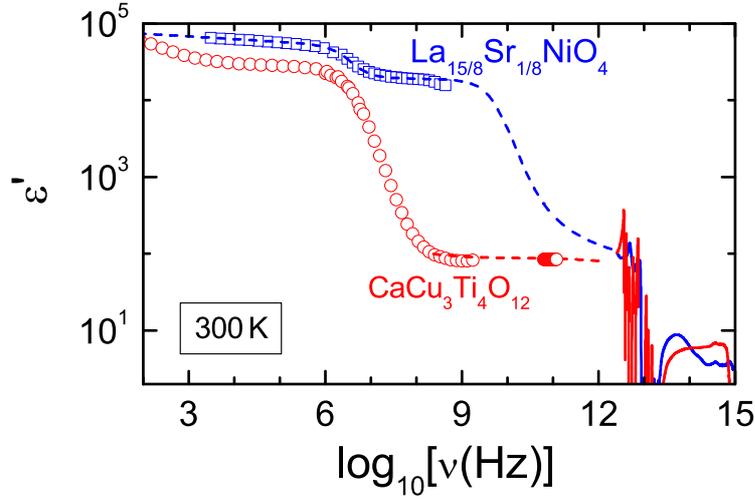} }}
\caption{Comparison of the frequency-dependent dielectric constant
of LSNO and CCTO single crystals from dielectric and IR measurements
\cite{Krohns2009b,Kant2008}. The measurements were performed at room
temperature. The upper dashed line was calculated from an
extrapolation of the parameters obtained from fits of spectra at
lower temperatures (cf. Fig. 2(a) of Ref. \cite{Krohns2009b}). The
lower dashed line is a guide to the eyes.} \label{fig:LSNO_optic}
\end{figure}

In Fig. \ref{fig:LSNO_optic}, a combined $\varepsilon'$-spectrum of
LSNO is shown for room temperature, including dielectric and
infrared data. For comparison also a corresponding spectrum for CCTO
is included (same data as in Fig. \ref{fig:CCTO_optic}
\cite{Kant2008}). This plot reveals a marked difference of the
dielectric response of both materials: While $\varepsilon'(\nu)$ of
CCTO starts to strongly decrease for frequencies beyond 1~MHz and is
far from being colossal at some 100~MHz and in the GHz region, this
is not the case for LSNO. Due to the restrictions of the
experimental technique employed to cover the MHz to GHz region when
measuring very high capacitances, no measurements in LSNO at
$\nu>430$~MHz were possible. However, the dashed line shows the most
likely course of $\varepsilon'(\nu)$ beyond this frequency. It is
based on an extrapolation of fit parameters from fits at lower
temperatures, where the MW-relaxation step is well within the
frequency window \cite{Krohns2009b,Krohns2010}. Overall, LSNO seems to be much
better suited than CCTO for applications at frequencies in the
MHz-GHz range, which is of high relevance, e.g., in modern
telecommunications technology.

Finally, we want to note that the infrared results in LSNO reveal a
"static" dielectric constant (which would be denoted as
$\varepsilon_{\infty}$ from a dielectric-spectroscopy viewpoint) of
approximately 90, much lower than the bulk value of 300 observed at
1~GHz (Fig. \ref{fig:LSNO_AuAg}(a)). This corroborates the
above-mentioned scenario of a dominating UDR contribution in
$\varepsilon'(\nu)$ up to the highest frequencies covered in the
dielectric measurements.

\begin{figure}[tbh]
\centerline{\resizebox{0.55\columnwidth}{!}{
  \includegraphics{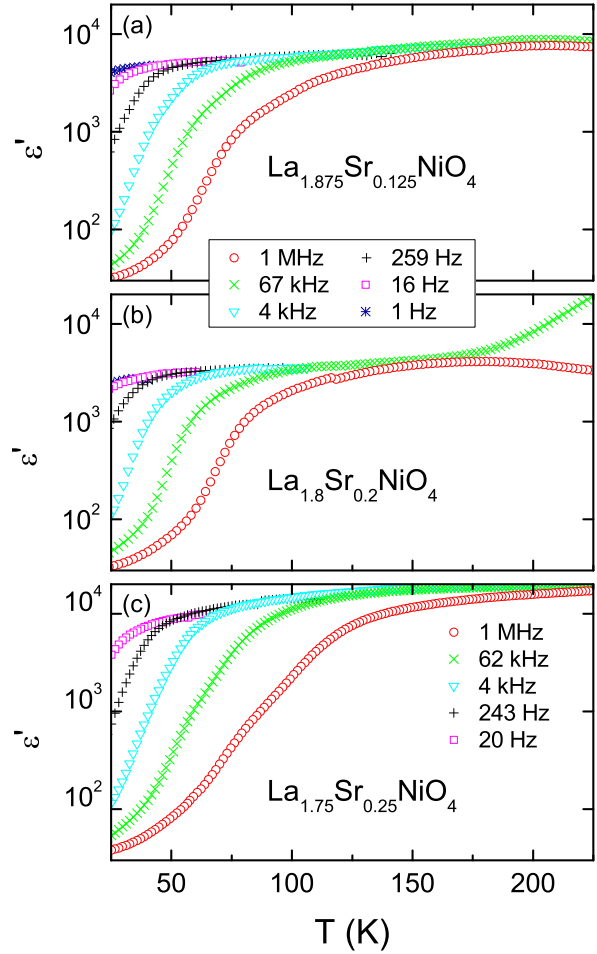} }}
\caption{Temperature dependent dielectric constant of
polycrystalline La$_{2-x}$Sr$_{x}$NiO$_{4}$ samples (tempered for
48~h at 1150~$^\circ$C, silver-paint contacts) with three different
Sr concentrations as measured at various frequencies.}
\label{fig:LSNO_xdep}
\end{figure}

The system La$_{2-x}$Sr$_{x}$NiO$_{4}$ provides the possibility of
tuning the Sr content over a wide range. Figure \ref{fig:LSNO_xdep}
shows first results demonstrating that, in addition to $x=1/8$, CDCs
are also found for other Sr contents, namely $x=0.2$ and 0.25 (see
also \cite{Rivas2004,Park2005,Liu2008}). In all cases the typical MW
relaxation is observed. The approximate agreement of the temperature
location of the relaxation steps for the poly- and
single-crystalline samples with $x=1/8$ (cf. Figs.
\ref{fig:LSNO_Tdep} and \ref{fig:LSNO_xdep}(a)) indicates that also
for the ceramic sample the CDC can be expected to persist up to
higher frequencies than in CCTO. The same can be said for the
compound with $x=0.2$ while for $x=0.25$ the relaxation steps seem
to be located at somewhat higher temperatures (i.e. the relaxation
time is higher). In addition, they are smeared out or even composed
of two separate relaxation steps and further investigations are
necessary to clarify this issue. In any case it is clear that
La$_{2-x}$Sr$_{x}$NiO$_{4}$ is a promising system that deserves at
least as much attention as CCTO and its relatives.

\subsection{Other transition-metal oxides with colossal dielectric constants}
\label{sec:others}

\begin{figure}[htb]
\centerline{\resizebox{1\columnwidth}{!}{
  \includegraphics{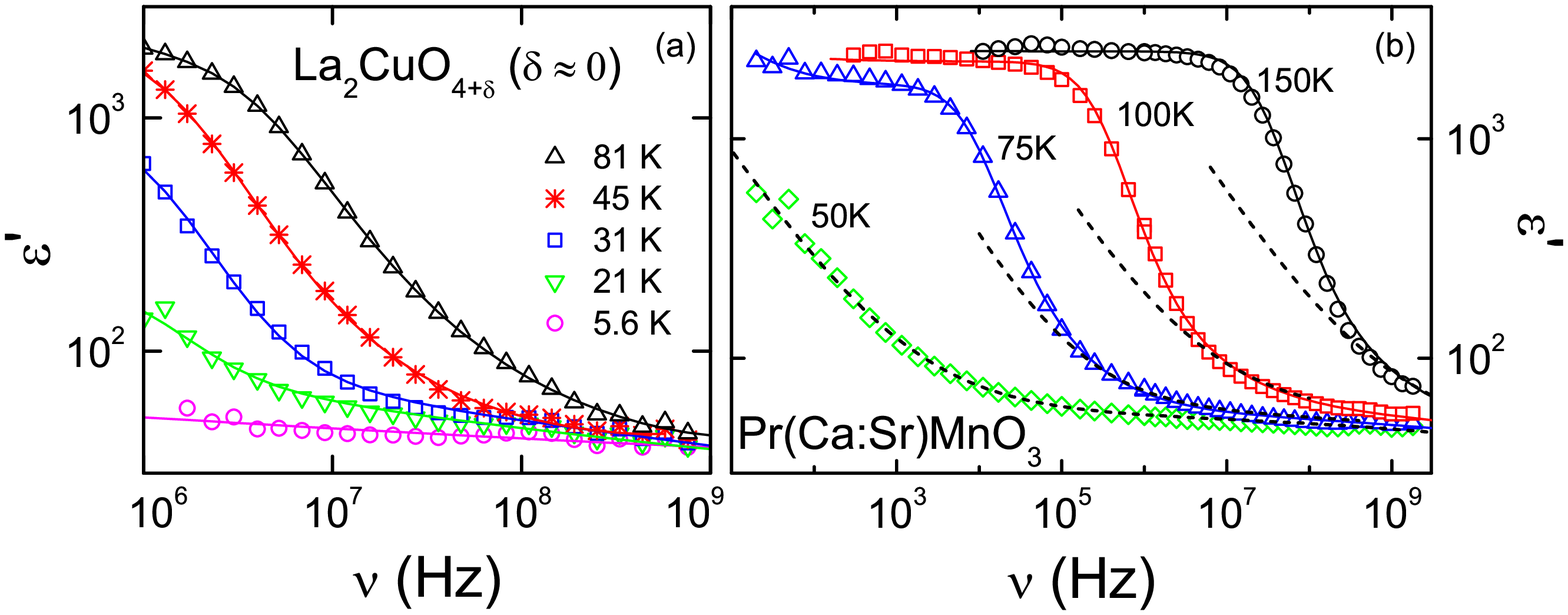} }}
\caption{Spectra of the dielectric constant measured at various
temperatures for La$_{2}$CuO$_{4}$
\cite{Lunkenheimer1992,Lunkenheimer1996} (a) and
Pr$_{0.65}$Ca$_{0.28}$Sr$_{0.07}$MnO$_{3}$ \cite{Sichelschmidt2001}
(b). The solid lines are fits with the equivalent circuit shown in
Fig. \ref{fig:theo} but without the second interface-related RC
element. The dashed lines in (b) show the intrinsic response.}
\label{fig:manyothers}
\end{figure}


As mentioned in the introductory section, there are many further
transition-metal oxides showing CDCs. Two typical examples are
provided in Fig. \ref{fig:manyothers}. It shows the frequency
dependence of the dielectric constant of single crystalline
$\rm{La_2CuO_{4+\delta}}$, measured at various temperatures. Details
have been reported in Refs.
\cite{Lunkenheimer1992,Lunkenheimer1996}. In this typical parent
compound of high-$T_c$ superconductors, the carrier concentration is
rather low $(\delta \sim 0)$. Nevertheless, at the bulk-electrode
interface a depletion layer is formed and the dielectric constant
reaches values of almost 2000 in the audio frequency range even at
low temperatures. As we were, at the time of this measurement, only
interested in intrinsic properties of $\rm{La_2CuO_{4+\delta}}$, we
focused on low temperatures and high frequencies $\nu>1$ MHz. At
elevated temperatures the contact contribution dominates the
response far into the microwave regime. The lines in Fig.
\ref{fig:manyothers}(a) are fits with the equivalent circuit shown
in Fig. \ref{fig:theo} but without the second interface-related RC
element. The intrinsic dielectric constant reaches values of
$\varepsilon_{\infty}\approx35$.

Figure \ref{fig:manyothers}(b) shows results on single-crystalline
Pr$_{0.65}$Ca$_{0.28}$Sr$_{0.07}$MnO$_{3}$ \cite{Sichelschmidt2001},
a colossal magnetoresistance material, which is very close to a
metal-insulator phase boundary and reveals antiferromagnetism and
charge order below about 200 K. Again the solid lines represent fits
using the equivalent circuit. The dashed lines indicate the
intrinsic response of the sample neglecting contact contributions.
The UDR leads to a $\omega^{s-1}$ contribution, which smears out the
contact-dominated step in $\varepsilon'(\omega)$ at high
frequencies. In  $\rm{Pr(Ca:Sr)MnO_3}$  the  intrinsic dielectric
constant $\varepsilon_{\infty}=50$.

\begin{figure}[htb]
\centerline{\resizebox{0.65\columnwidth}{!}{
  \includegraphics{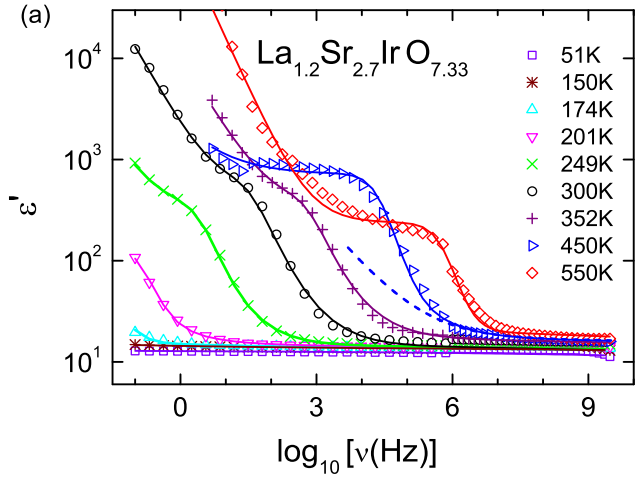} }}
\caption{Spectra of the dielectric constant of
La$_{1.2}$Sr$_{2.7}$IrO$_{7.33}$ for various temperatures
\cite{Lunkenheimer2006}. The solid lines are fits with the
equivalent circuit shown in Fig. \ref{fig:theo} but without the
second interface-related RC element. The dashed line shows the
intrinsic response for 450~K.} \label{fig:LSIO}
\end{figure}

Finally, Fig. \ref{fig:LSIO} provides the dielectric constant of
La$_{1.2}$Sr$_{2.7}$IrO$_{7.33}$ \cite{Lunkenheimer2006}. This
compound is formed by alternative stacking of hexagonal perovskite
($A_{2}$IrO$_{6}$) and $A'_{2}$O$_{1+\delta}$ layers ($A =$~La/Sr;
$A' =$~Sr) \cite{Gotzfried2005}. In this material, both oxygen and
peroxide ions are present, occupying large cavities formed by six
$A$O$_{6}$ prisms. Within these cavities, the ions can occupy six
different off-centre positions and also the $A'$ ions can assume
three different positions within the $A'_{2}$O$_{1+\delta}$ layers.
Ions in off-centre positions can generate ferroelectricity and also
may lead to dipolar relaxation phenomena, which makes this material
interesting from a dielectric-spectroscopy viewpoint. In addition,
the strong substitutional disorder should give rise to
charge-carrier localisation and the typical signatures of this
phenomenon in the ac response. As seen in Fig. \ref{fig:LSIO}, for
low frequencies $\varepsilon'(\nu)$ reaches the typical plateau of
the MW relaxation with CDCs of the order of 1000. However, for
further decreasing frequency an additional strong increase shows up.
It exhibits no indication of a second plateau as would be expected
for a second relaxation feature. Fits of these spectra, which were
also simultaneously performed for the conductivity
\cite{Lunkenheimer2006}, were able to cover this extra increase
without assuming a second relaxation (i.e. without a second
interfacial RC element, as shown in the equivalent circuit of Fig.
\ref{fig:theo}(c)). A detailed analysis revealed that this
low-frequency increase of $\varepsilon'(\nu)$ is due to the strong
intrinsic UDR contribution in this material (indicated by the dashed
line for 450~K). It is sufficiently strong to also show up at
frequencies below the CDC plateau of the MW relaxation and thus this
material provides a nice example for CDCs generated by hopping
charge transport (see Sec. \ref{sec:hop}). Measurements with
different contact types in this material \cite{Lunkenheimer2006}
indicate that the MW relaxation is due to an SLBC mechanism but even
without this effect, very large values of $\varepsilon'$ would be
reached at low frequencies. For a detailed discussion of the
dielectric properties of La$_{1.2}$Sr$_{2.7}$IrO$_{7.33}$, also
revealing an intrinsic dipolar relaxation, see Ref.
\cite{Lunkenheimer2006}.


\section{Summary and Conclusions}
\label{sec:concl}

A large variety of physical mechanisms can give rise to a colossal
magnitude of the dielectric constant. This includes
ferroelectricity, with all its disadvantages for technical
application, charge-density-wave formation and the approach of a
metal-insulator transition, which mostly are of high academic
interest only, and interfacial polarisation with its many subgroups
as SBLC and IBLC generated effects, including, e.g., electronic
phase separation. As revealed by the results provided in the present
work, transition-metal oxides in general seem to be prone to the
occurrence of colossal magnitudes of the dielectric constant. It is
clear that in many cases interfacial barriers within or at the
surface of the samples play an important role in the generation of
the observed CDCs. Especially in the extensively investigated CCTO,
despite some sophisticated attempts of invoking intrinsic
mechanisms, a non-intrinsic MW process seems the most likely
explanation of its CDCs. However, despite epic efforts of innumerous
groups during the past ten years, no consensus on the nature of the
involved interfaces has been achieved. In light of the results
presented in this work it is clear that in CCTO and also in most
other materials, thin insulating layers at the sample surface, most
likely induced by diode formation between the metallic electrodes
and the semiconducting sample, must play a role. However, also
indications for IBLCs of various origins were found. It seems that
the question "What causes the CDCs in CCTO?" cannot be unequivocally
answered and following the line proposed in Ref.
\cite{Ferrarelli2009} the answer may depend on the specific sample
investigated and also combinations of different effects seem likely.

Another outcome of the present work is the finding that CCTO is only
one member of a large group of isostructural materials with similar
properties, examples being La$_{2/3}$Cu$_{3}$Ti$_{4}$O$_{12}$ and
Pr$_{2/3}$Cu$_{3}$Ti$_{4}$O$_{12}$. Those material show promising
dielectric properties and one can expect to discover other, even
better suited ones within this vast group of compounds, that indeed
once may become new standard materials for high-capacitance
applications.

Irrespective of any application viewpoints one should not forget
that CCTO and its relatives are highly interesting materials also
for purely academic reasons. For example, its relatively high
intrinsic dielectric constant, further increasing when lowering the
temperature, the softening of several of its phonon modes and the
occurrence of isosbestic points in the infrared reflectivity spectra
\cite{Kant2008} deserves further investigation. It seems likely that
ferroelectric correlations are present in CCTO and based on its
crystal structure, the Ti$^{4+}$ cation, rattling within the
TiO$_{6}$ octaeder and tending to go off-centre at low temperatures
\cite{Subramanian2000} seems a likely mechanism. Even incipient
ferroelectricity may be possible \cite{Li2007}.

Among the many other transition-metal oxides with CDCs, the system
La$_{2-x}$Sr$_{x}$NiO$_{4}$ stands out by showing electronic phase
separation that may well play an important role in the generation of
the observed CDCs in this material. Of special significance is our
finding that at room temperature $\varepsilon'(\nu)$ of this
material remains colossal well up to the GHz frequency range, quite
in contrast to CCTO.

\medskip
  {\small
This work was supported by the DFG via the SFB 484 and by the
Commission of the European Communities, STREP: NUOTO,
NMP3-CT-2006-032644.}

%
%

\end{document}